\documentclass[11pt,a4paper]{article}
\usepackage[margin=1in]{geometry}

\usepackage[natbibapa]{apacite}
\usepackage{hyperref}
\usepackage{amsmath}
\usepackage{graphicx}
\usepackage{esvect}
\usepackage{amssymb}
\numberwithin{equation}{section}
\usepackage{float}
\usepackage{todonotes}
\usepackage{tabularx}
\usepackage{booktabs}
\usepackage{setspace}
\usepackage[ruled,vlined]{algorithm2e}
\usepackage{float}
\usepackage{longtable}  

\usepackage{comment}

\usepackage{caption} 
\usepackage{array} 
\newcolumntype{P}[1]{>{\centering\arraybackslash}p{#1}}
\newcolumntype{M}[1]{>{\centering\arraybackslash}m{#1}}

\usepackage[utf8]{inputenc}
\usepackage[english]{babel}

\graphicspath{ {./images/} }

\title{Solving the unit-load pre-marshalling problem in block stacking storage systems with multiple access directions\thanks{This work was partly funded by the Deutsche Forschungsgemeinschaft (DFG, German Research Foundation) through the Research Training Group 2193 (Project number: 276879186).}}

\author{Jakob Pfrommer, Anne Meyer, Kevin Tierney}
\author{ Jakob Pfrommer\footnote{Corresponding author, TU Dortmund University, Leonhard-Euler-Stra{\ss}e 5, 44227 Dortmund, Germany, jakob.pfrommerk@tu-dortmund.de, ORCID: \url{https://orcid.org/0000-0003-2492-0621}}, Anne Meyer\footnote{TU Dortmund University, Leonhard-Euler-Stra{\ss}e 5, 44227 Dortmund, Germany, anne2.meyer@tu-dortmund.de, ORCID: \url{https://orcid.org/0000-0001-6380-1348}}, Kevin Tierney\footnote{Bielefeld University, Universitaetsstra{\ss}e 25, 33615 Bielefeld, Germany, kevin.tierney@uni-bielefeld.de, ORCID: \url{https://orcid.org/0000-0002-5931-4907}}}
\date{}

\begin{document}
\maketitle

\begin{abstract}
Block stacking storage systems are highly adaptable warehouse systems with low investment costs. 
With multiple, deep lanes they can achieve high storage densities, but accessing some unit loads can be time-consuming. 
The unit-load pre-marshalling problem sorts the unit loads in a block stacking storage system in off-peak time periods to prepare for upcoming orders. The goal is to find a minimum number of unit-load moves needed to sequence a storage bay in ascending order based on the retrieval priority group of each unit load. 
In this paper, we present two solution approaches for determining the minimum number of unit-load moves. We show that for storage bays with one access direction, it is possible to adapt existing, optimal tree search procedures and lower bound heuristics from the container pre-marshalling problem. For multiple access directions, we develop a novel, two-step solution approach based on a network flow model and an A* algorithm with an adapted lower bound that is applicable in all scenarios. 
We further analyze the performance of the presented solutions in computational experiments for randomly generated problem instances and show that multiple access directions greatly reduce both the total access time of unit loads and the required sorting effort.

\end{abstract}

\section{Introduction}
\label{1_intro}
Warehouse systems are constantly adapted to dynamic material flows due to changing customer demand or evolving production capacities. 
Block stacking storage systems are especially capable of dealing with these challenges thanks to their flexibility, expandability and ease of setup/breakdown. 
In block stacking storage systems, unit loads (e.g., pallets or boxes) are placed on the floor and stacked on top of each other in a grid-based layout. 
Block storage does not require the installation of technical infrastructure. This keeps investment costs low and allows companies to quickly react to changing business conditions. Types of block storage systems consisting of movable units on the floor include, for example, pallet-based block storage, robotic mobile fulfillment systems (RMFS), container yards or buffer storage with material transport trolleys.

\begin{figure}[!ht]
    \centering
    \includegraphics[width=9cm]{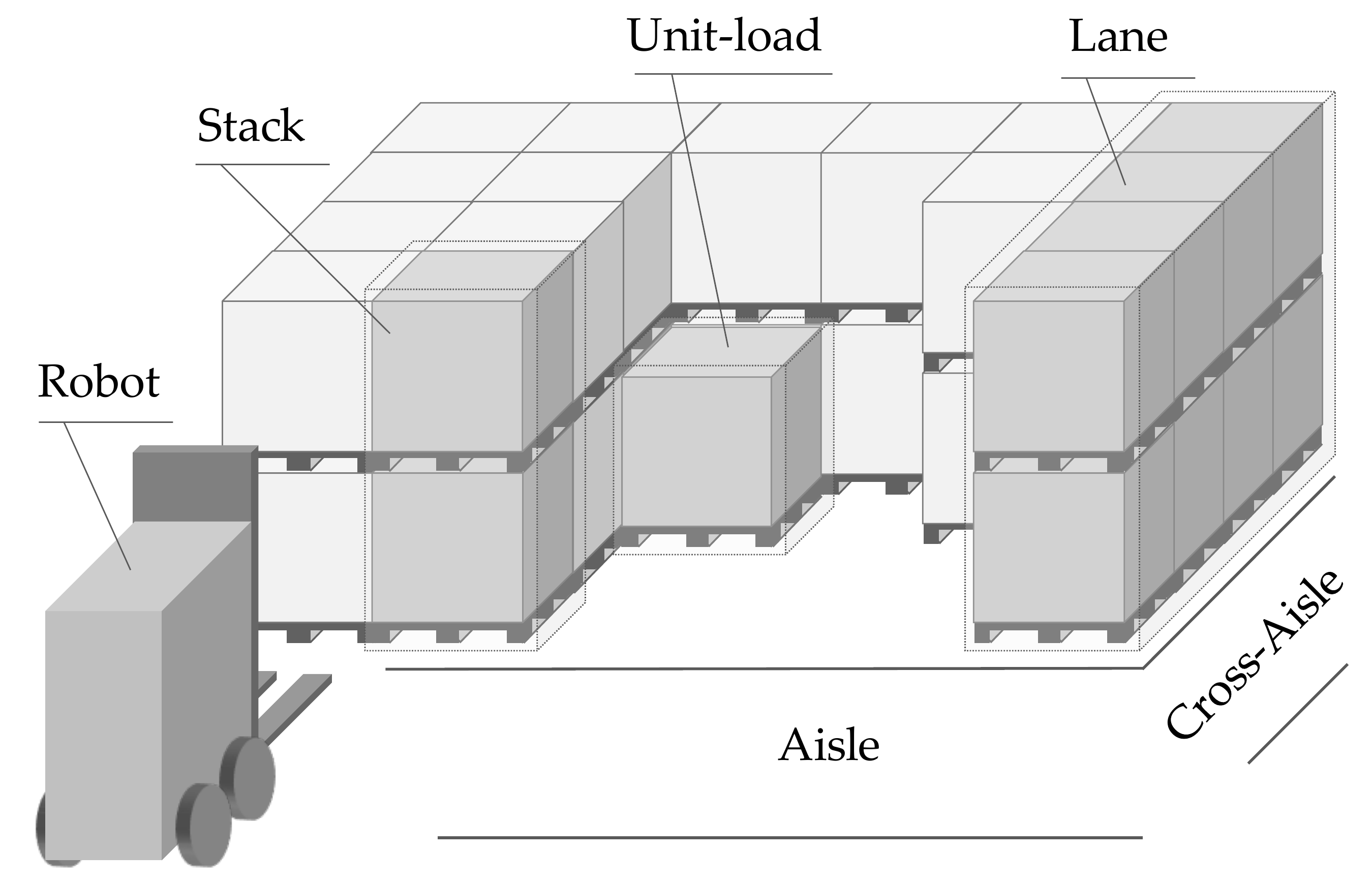}
    \caption{Example of a single storage bay in a block stacking storage system with a robotic forklift.}
    \label{fig:block_storage}
\end{figure}

Figure \ref{fig:block_storage} shows the basic structure of a block stacking storage system. Each lane of a storage bay is filled with a sequence of stacks. Only the outermost stacks can be accessed by transport vehicles like robotic forklifts. In this type of storage system, unit loads can be accessed from multiple directions, i.e., from an adjacent aisle or cross-aisle. This provides immense flexibility compared to systems that can only be accessed from the front. However, unit loads can still be blocked from being retrieved if other unit loads are (i) stored on top of them or (ii) stored between the unit load and the direction of access (aisle/cross-aisle), and the blocking unit load has a later retrieval time than the one being removed.

Autonomous mobile robots (AMRs) (more specifically in this work, robotic forklifts) 
are being increasingly used to control material handling and storage systems \citep{fragapane2021planning}. However, most block stacking warehouses are currently still organized manually. Human operators generally compose warehouses with a fixed layout and storage zone structure, 
in which stock keeping units (SKUs) are always placed at defined storage locations (e.g., a single SKU is always assigned the same lane).

In autonomously organized block stacking warehouses, the advantage of robots is that they have exact knowledge of the system state, including the storage location 
of each unit load. Thus, they naturally support storage policies without a fixed allocation of SKUs to storage locations (shared storage policy) as well as extensive rearrangement 
operations. 
This increases the storage density by better utilizing the available storage space. 
Another advantage of AMRs is that they are operational around the clock. Hence, off-peak hours can be utilized to sort the system, improving future access and retrieval times. For example, the storage system could prepare for an expected demand sequence like the loading times of trucks or the production schedule over the next few days. We introduce this problem for block stacking storage systems as the unit-load pre-marshalling problem (UPMP). An overview of major decision problems in autonomously organized block stacking warehouse is provided in \citet{pfrommer2020autonomously}.

For the UPMP, 
we assume that the current warehouse configuration with the position and retrieval priority group for each unit load is given and that no unit loads enter or leave the warehouse during planning/rearranging. The allocation of unit loads to a corresponding retrieval priority group is based on a known or predicted future retrieval time 
(see also the duration-of-stay storage strategy \cite{goetschalckx1990shared}).
The goal is to find a minimum number of unit-load moves needed to sort 
a storage bay such that all unit loads can be retrieved according to their retrieval group without any blockage. Unit loads can only be accessed by AMRs at the outermost stacks if a free path from at least one of the four cardinal directions is available. Direct access may be blocked by further unit loads, walls or other obstacles, as well as due to limitations of the load carriers. 

\begin{figure}[!ht]
    \centering
    \includegraphics[width=12cm]{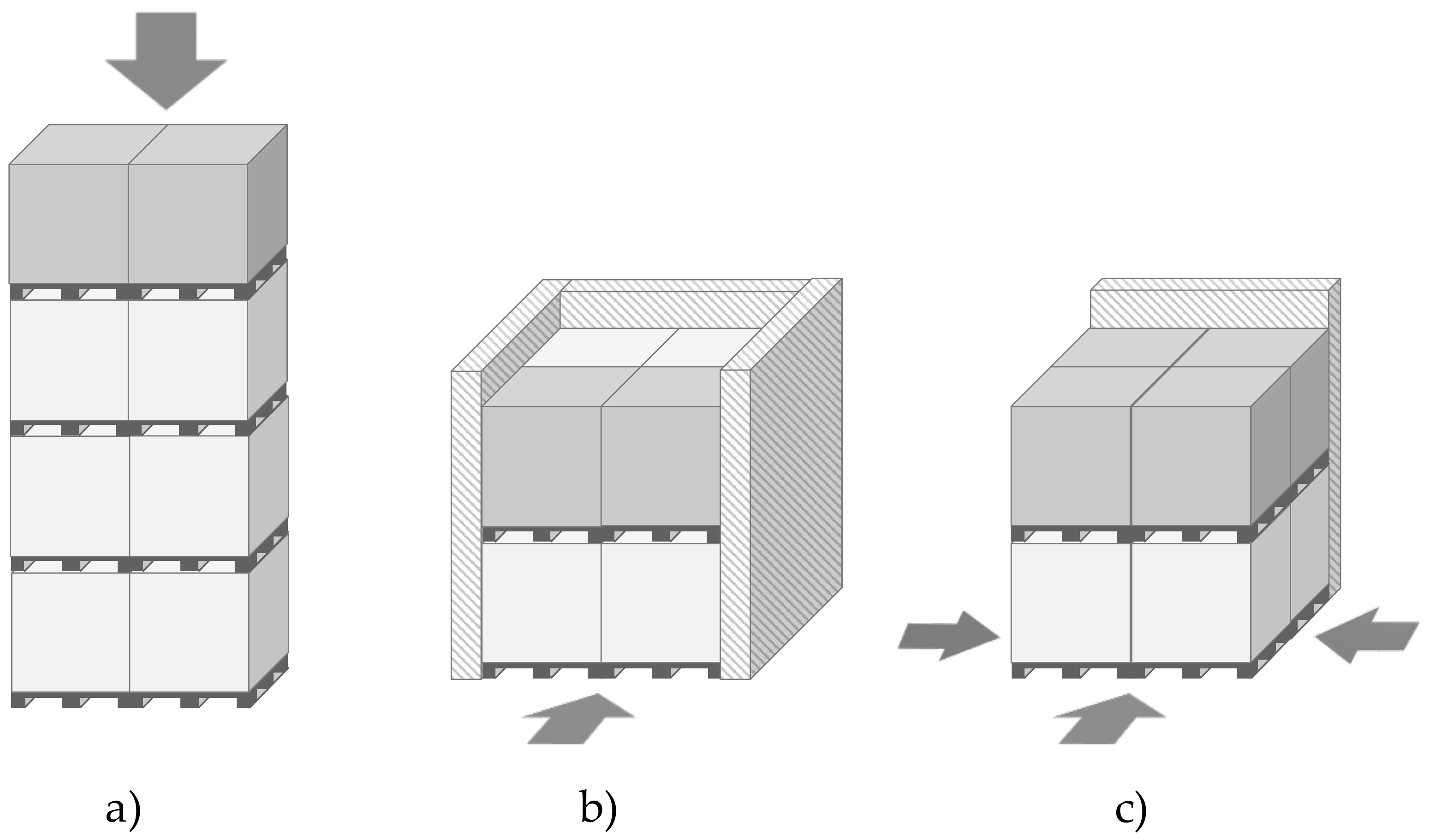}
    \caption{Example storage bays with eight unit loads in the CPMP via crane in a) as well as side access in the UPMP via one direction in b) and three directions in c). The directly accessible unit loads are dark-gray.}  
    \label{fig:access_outermost_stacks}
\end{figure}

The pre-marshalling problem is a well-known problem in the container stacking \citep{lehnfeldLoadingUnloadingPremarshalling2014, casertaContainerRehandlingMaritime2011} and steel slab and coil shuffling literature \citep{tangModelsAlgorithmsShuffling2012, geLogisticsOptimisationSlab2020}. In the container pre-marshalling problem (CPMP), all stacks are accessible from the top via a crane. Figure \ref{fig:access_outermost_stacks} a) illustrates an example with two stacks containing four unit loads. The UPMP with side access from only one direction in Figure \ref{fig:access_outermost_stacks} b) is similar to the CPMP. In a sequence of stacks per lane, the unit loads at the outermost stacks can be accessed directly. Both examples \ref{fig:access_outermost_stacks} a) and \ref{fig:access_outermost_stacks} b) allow direct access to only two of the eight unit loads (dark-gray). To reach all other unit loads, relocation of the blocking unit loads on top or in front is necessary. 
A common setup in warehouses is the use of pallets and forklift trucks, which allow access from at least two opposite directions, in many cases even from four directions. Figure \ref{fig:access_outermost_stacks} c) shows an example of the UPMP, where three access directions are available and only the back is blocked by a wall. This leads to direct access to every stack, which speeds up access times and reduces the number of reshuffling moves necessary to sort the warehouse. To our knowledge, a stacking problem with side access to bays from up to four access directions has not been studied yet.

In this paper, we formally introduce the UPMP and propose solution approaches for finding a minimum number of unit-load moves to sort a block stacking storage system. We show that for the UPMP 
with one access direction, it is possible to transfer existing solution approaches from the CPMP. For cases with multiple access directions, 
we develop a novel two-step solution approach based on a network flow model and an adapted search procedure that is applicable in all scenarios with up to four access directions. The proposed tree search methods and lower bound heuristics based on algorithms from the CPMP (see \cite{tierneySolvingPremarshallingProblem2017}) solve the UPMP to optimality. Finally, our contribution is complemented by a publicly available benchmark set and experimental evaluation of the proposed tree search methods. 

This paper is organized as follows. We first introduce 
the UPMP in Section \ref{2_UPMP}. This is followed by a literature overview 
in Section \ref{3_related_work}. Sections \ref{4_UPMP_one} and \ref{5_UPMP_two} describe our solution approaches for one and up to four access directions, respectively.
In Section \ref{6_experiments}, we present our UPMP benchmark dataset and 
provide the results of our computational experiments. 
The impact of multiple access directions are discussed in Section \ref{7_insights}. Finally, we conclude and point out future research directions 
in Section \ref{8_conlusion}.

\section{The unit-load pre-marshalling problem} \label{2_UPMP}
The UPMP involves sorting a bay of a block stacking storage warehouse until all blockage is resolved. We assume that no items are going in or out of the warehouse, i.e., the AMRs are idle. The given bay is fully defined by a set of unit loads and their retrieval groups under a fully shared storage strategy (SKUs are mixed in stacks and lanes). 
Blockage occurs if unit loads of a higher retrieval group are placed in front of a unit load of a lower retrieval group. We define such unit loads as \textit{blocking}. 
The goal is to find a minimum number of unit-load moves to sort 
a storage bay such that there is no blockage in the bay. The resulting sequence of moves can be executed by one or multiple vehicles. 

For the UPMP, we make the following assumptions with respect to the technical features 
of the block stacking warehouse. Since we only consider a single storage bay, the minimization of the number of moves is reasonable as an approximation of the sorting effort. We assume that 
process times for positioning and handling 
of AMRs dominate driving times. 
For larger warehouse setups with multiple bays, it would also be necessary to take into account travel times. Furthermore, we assume that robots can only carry a single unit load at a time and cannot reach over the top of a stack in front. 
%
If a stack has multiple tiers, only the top unit load can be accessed directly. This corresponds to the capabilities of standard industrial forklift trucks. Increasing the transport capacity to multiple items is a possible modification. 

\begin{figure}[!ht]
    \centering
    \includegraphics[width=8.5cm]{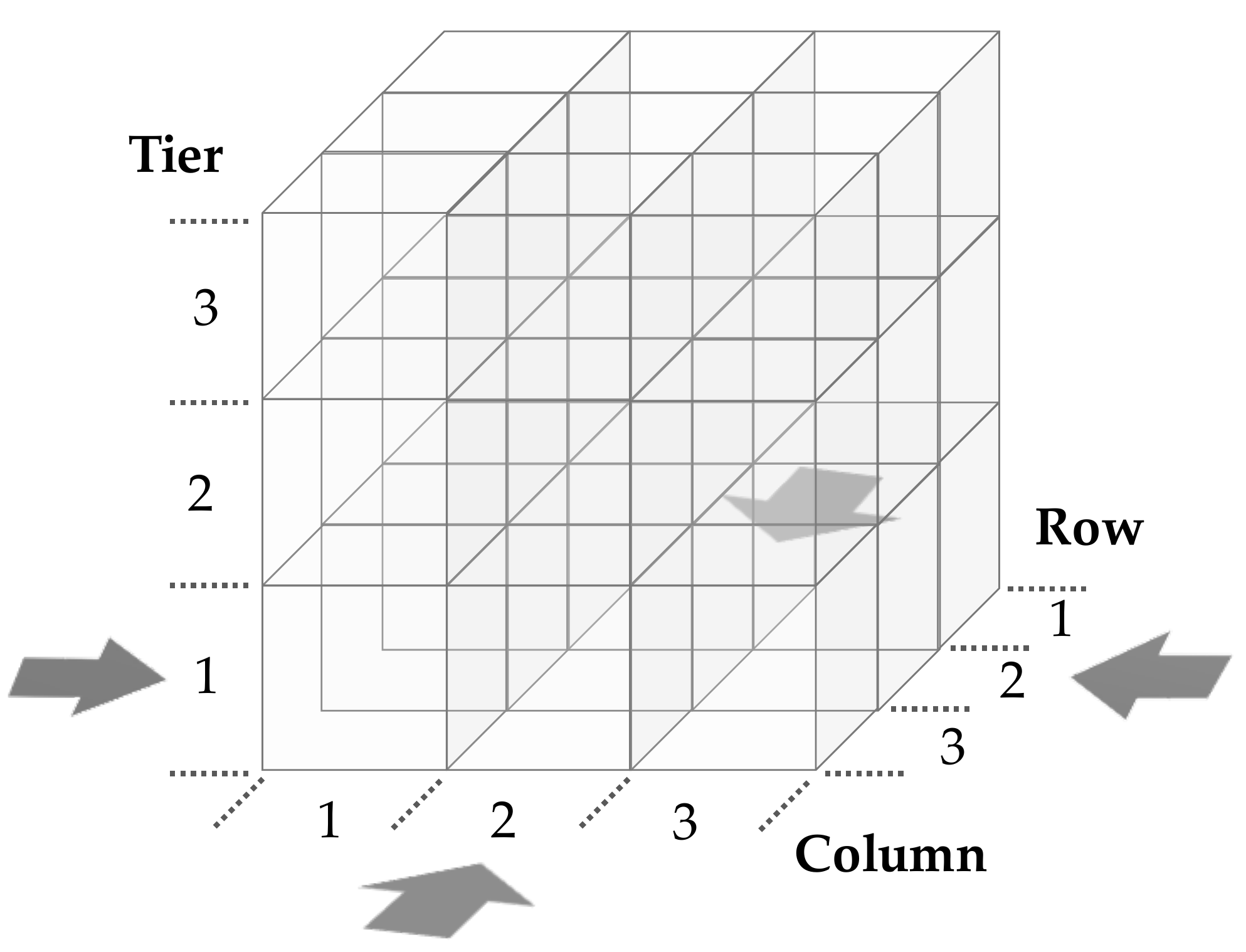}
    \caption{State representation of the UPMP.}
    \label{fig:State_rep}
\end{figure}

As shown in Figure \ref{fig:State_rep}, a storage bay in the UPMP consists of a three-dimensional grid of storage locations with columns $I$, rows $J$ and tiers $T$. Unit loads can only be placed above tier one if unit loads are stored underneath. Also, bays are filled up and sorted dependent on the access direction, preventing empty space or gaps in between.

\begin{figure}[!ht]
    \centering
    \includegraphics[width=\textwidth]{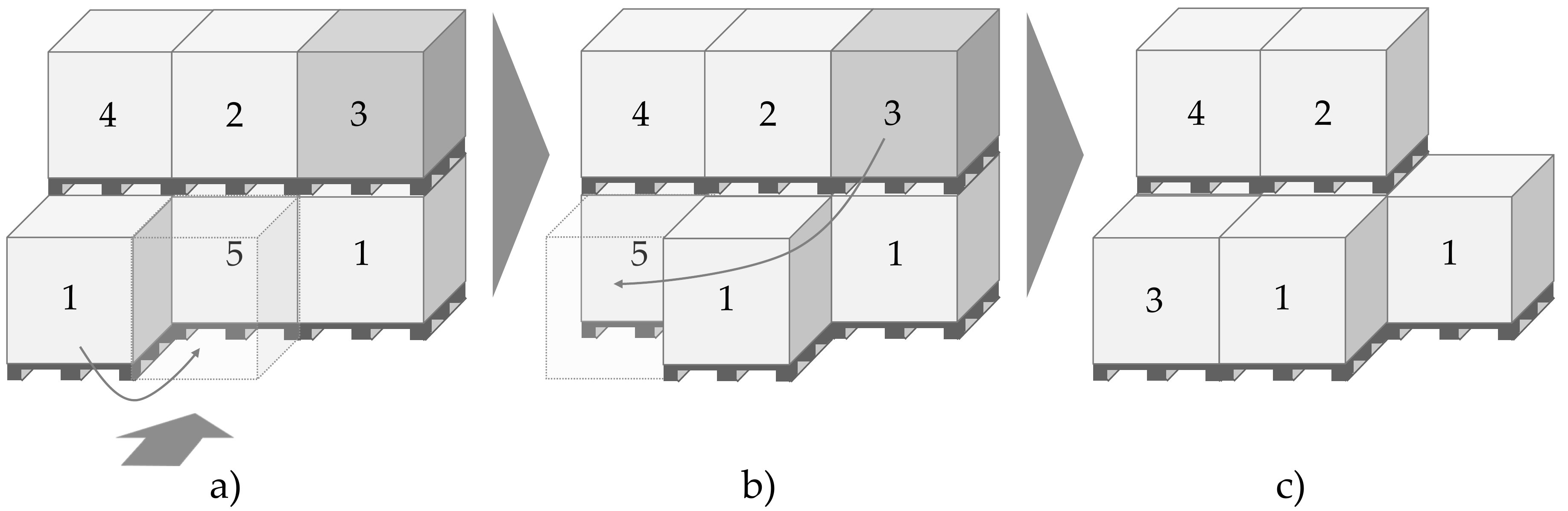}
    \caption{Example configuration and solution of the UPMP}
    \label{fig:Sort_example}
\end{figure}

Figure \ref{fig:Sort_example} shows a simple example of a configuration with $I=3$, $J=2$ and $T=2$ along with a possible solution of the UPMP with one access direction. Retrieval groups $g_{ijt} \in\{1,...,G\}$ from one to five are assigned to each unit load in the storage bay at their respective position ($i \in\{1,...,I\}$, $j \in\{1,...,J\}$, $t \in\{1,...,T\}$). The initial configuration in Figure \ref{fig:Sort_example} a) is sorted according to the retrieval groups by a sequence of two consecutive moves. The unit load of group $3$ cannot be moved in front of the groups $1$ and $2$ of the other two lanes. So in a first move, we need to free up a storage location for the blocking unit load of group $3$ and reach the new configuration in Figure \ref{fig:Sort_example} b). A second move leads to Figure \ref{fig:Sort_example} c). This is the final, sorted configuration of the bay. 

The main idea of unit-load pre-marshalling is to increase the storage density of storage systems by sorting them in idle time so that time-consuming relocation moves can be avoided later on in peak hours. 
The UPMP is an operational problem, which may arise several times throughout the day, overnight or less frequently (e.g., on the weekend).
Given the proven complexity of the CPMP, which is NP-hard (see \cite{casertaContainerRehandlingMaritime2011}), the similar UPMP with one access direction is also NP-hard. The cases with more than one access direction are a generalization of the problem and thus also NP-hard. 

\section{Related work} 
\label{3_related_work}
Relocation problems for sorting a storage system in idle time have been widely studied in many different domains. Due to their domain-specific nature, we categorize the related decision problems by the area of application, namely container terminals, the steel industry, block stacking warehouses, puzzle-based storage systems, robotic mobile fulfillment systems, and automated storage and retrieval systems (AS/RS). 

\subsection{Container terminals}
The CPMP is a well-studied problem concerning the reshuffling of a set of containers in a single storage bay using an overhead crane. A variety of exact and heuristics methods have been used to solve the CPMP. 
Methods for solving the CPMP include integer programming \citep{lee2007optimization,parreno-torresIntegerProgrammingModels2019}, dynamic programming \citep{prandtstetter2013dynamic}, constraint programming \citep{rendlConstraintModelsContainer2013} and several tree search approaches, which we briefly present in the following as they are the foundation of this paper.

Tree-based approaches using the A* algorithm~\citep{hart1968formal} have been successful at solving real-world sized CPMP instances to optimality. The three primary ingredients of these approaches are (i) the branching heuristic, (ii) symmetry breaking rules and (iii) the lower bound heuristic.
\citet{exposito-izquierdoPreMarshallingProblemHeuristic2012} introduce an A* approach for the CPMP using a simple lower bound that only counts blocking items. \citet{bortfeldtTreeSearchProcedure2012} describe a heuristic tree search procedure that uses a lower bound based on the ideas of \emph{supply} (available storage locations) and \emph{demand} (available storage locations) for each retrieval priority group. \citet{tierneySolvingPremarshallingProblem2017}, \citet{tanakaSolvingRealworldSized2018} and \citet{tanakaBranchBoundApproach2019} build on this bound and tighten it within an iterative deepening search to reduce memory consumption. Furthermore, \citet{tierneySolvingPremarshallingProblem2017} and ~\cite{tanakaBranchBoundApproach2019} propose several branching and symmetry breaking rules. 

In intra-block re-marshalling, the problem scope is expanded beyond a single bay, allowing containers to move between multiple bays. This problem is not yet well studied, partially because an assumption of the CPMP is that moving the crane between bays is not possible or costly for pre-marshalling operations. We refer to the review of \citet{casertaContainerRehandlingMaritime2011} and the publication of \citet{covicRemarshallingAutomatedContainer2017}, who compares different features of existing re-marshalling algorithms with his own contribution. 

\subsection{Steel industry}
Shuffling problems in the steel industry also bear resemblance to the UPMP. \citet{tangModelsAlgorithmsShuffling2012} categorizes these problems into plate shuffling problems (e.g., \citet{tangModellingGeneticAlgorithm2002, konigSolutionsRealWorldInstances2007}) and coil shuffling problems (e.g., \citet{Zpfel2006WarehouseSI, tangModelsAlgorithmsShuffling2012, tangModelingSolutionShip2015}). The slab pre-marshalling problem (SPMP) introduced by \citet{geLogisticsOptimisationSlab2020} deals with sorting steel slabs during idle times. Unlike the approach of the CPMP where items are sorted according to their retrieval sequence, the goal in the SPMP is to generate stacks consisting of a single item group. 
The integrated optimization of storage and pre-marshalling moves by \citet{geIntegratedOptimisationStorage2021a} combines the SPMP and storage decisions to minimize blocking items in as few stacks as possible.

\subsection{Block stacking warehouses}
In terms of block stacking warehouses, only the recent publication from \citet{maniezzoStochasticPremarshallingBlock2021} is concerned with pre-marshalling. They analyze the stochastic pre-marshalling problem, where a future retrieval sequence is not exactly known. Based on historical data, a statistical model is used to derive a probability distribution for upcoming requests of each SKU. This is applied to estimate the number of moves in their pre-marshalling algorithms. In contrast to this work, all stacks are directly accessible. Hence, only blockage within a stack is considered.

\subsection{Puzzle-based storage systems} 
Stacking problems are similar to the well-known 8 and 15 puzzle problems in artificial intelligence. In \cite{ballMathematicalRecreationsEssays1893}, puzzle-based storage systems items are placed in a highly dense grid-based layout. In this layout, at least one of the locations in the grid must be empty to enable items retrieval. 
Items are shuffled between the empty location(s) until a desired goal configuration is reached. 
\citet{guePuzzlebasedStorageSystems2007} provide a comparison of puzzle-based and aisle-based storage. The analysis of different configurations shows that the moving costs and expected retrieval time are higher for puzzle-based storage than for classic aisle-based layouts.

\subsection{RMFS}
An RMFS being increasingly used in practice involves mobile racks (pods) that are moved by AMRs, and is especially popular in e-commerce fulfillment centers (e.g., Kiva systems \citet{wurman2008coordinating}). \citet{merschformannActiveRepositioningStorage2018} differentiates between active and passive repositioning in such systems. Active repositioning involves actively starting a relocation process, similar to the UPMP when conducted in idle time. Passive repositioning, in contrast, consists of updating a storage position when pods are returning from a picking station. 
Blockage is relevant in RMFS with multi-deep storage bays as shown by \citet{jinMultipleDeepLayout2020a} and \citet{yangModellingAnalysisMultideep2021}. The authors point out that additional working time is required to relocate obstructing pods.

\subsection{AS/RS} 
Another stream of publications of relocation problems is related to AS/RS. Pre-marshalling in terms of sorting a warehouse to resolve blockage is only relevant for multi-deep storage setups 
(e.g.,  AutoStore \citet{zouEvaluatingDedicatedShared2016}). However, in the literature, pre-marshalling problems for these multi-deep systems have hardly been formulated. An investigation of the storage location assignment problem with regard to blockage in deep-lane storage for one and two opposite access directions is presented by \citet{boysen2018deep}. 
\section{Solving the UPMP with a single access direction} 
\label{4_UPMP_one}
We present a search procedure based on the A* algorithm to solve the UPMP with one allowed access direction to optimality, analogous to the approaches presented by \citet{exposito-izquierdoPreMarshallingProblemHeuristic2012} and \citet{tierneySolvingPremarshallingProblem2017} for the CPMP. 
We first describe the A* algorithm  implemented branching rules, followed by the adapted lower bound. 

\subsection{A* for the unit-load pre-marshalling problem}
The 
A* algorithm 
uses a best first search guided by a heuristic to solve planning problems.
Starting from an initial bay configuration all possible moves (branches) of each movable unit load are examined. 
Iteratively the most promising move that minimizes costs is extended. 
The cost estimation heuristic for branch selection 
is given by $f(n) = g(n) + h(n)$, where $n$ represents a search node, $g(n)$ is the number of performed moves at node $n$, and $h(n)$ is a lower bound on the number of moves needed to reach a configuration from node $n$ without any blocked unit loads. 

\begin{figure}[!ht]
    \centering
    \includegraphics[width=11.5cm]{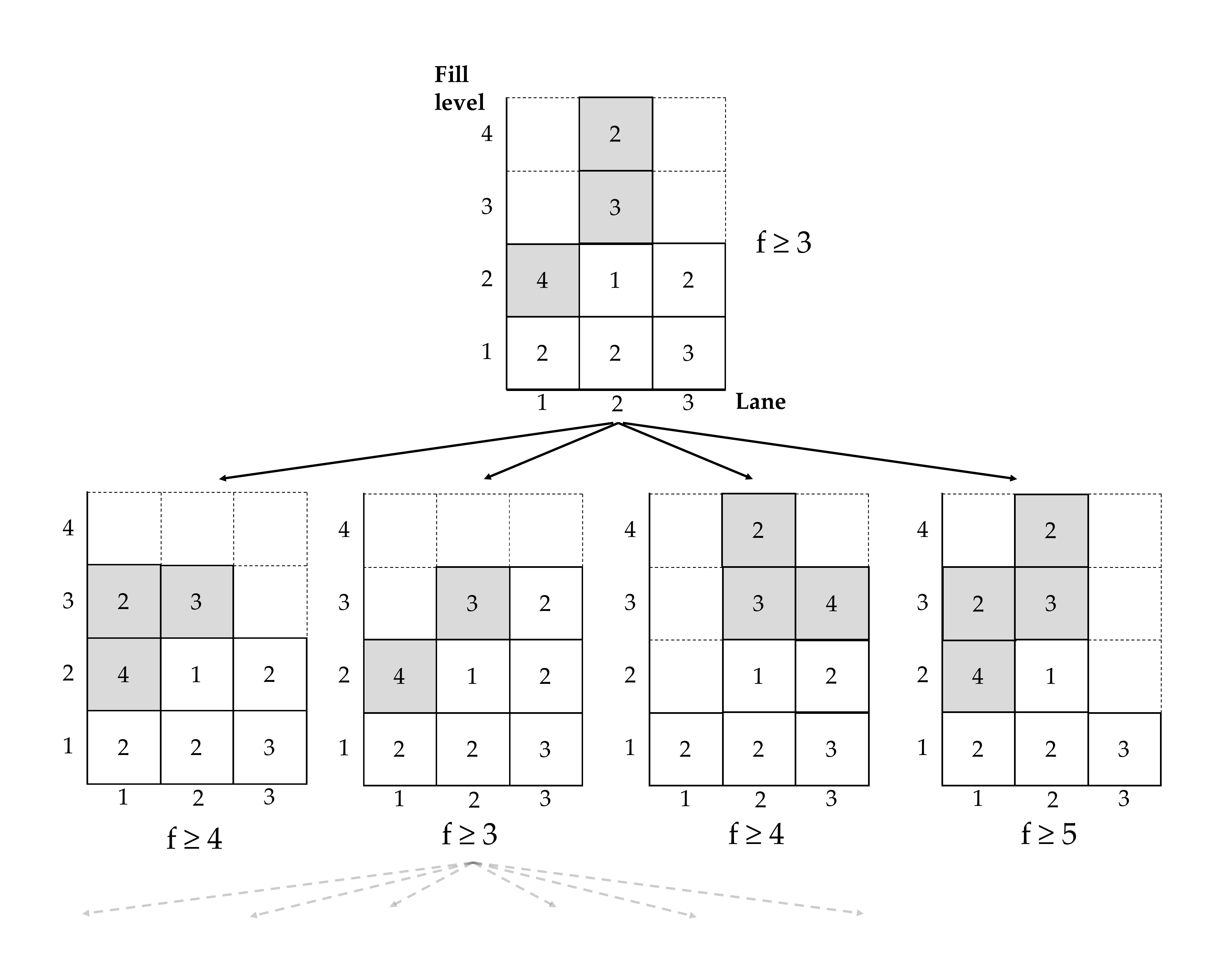}
    \caption{Example of the tree search procedure (simple lower bound as blocking items in dark gray) similar to the illustration in \citet{tierneySolvingPremarshallingProblem2017}}
    \label{fig:Example_tree_search}
\end{figure}

Figure \ref{fig:Example_tree_search} illustrates the search procedure with an example based on counting blocking items as a simple lower bound. Starting from an initial configuration with three blocking items, there are four possible moves, leading to four nodes all with a search depth of one. These newly generated nodes are called \emph{open nodes}, and we calculate $f(n)$ for each one. We explore the node with the lowest value of $f(n)$, which in this example is the configuration with $f(n)=3$. We apply tie breaking rules and symmetry breaking that is described in more detail later. 

Algorithm \ref{astar} shows the A* algorithm for the UPMP \citep{tierneySolvingPremarshallingProblem2017}. It starts with the initialization of an empty open list where all unexplored nodes are stored, an open dictionary mapping the unexplored nodes to f-values and a closed set collecting all previously visited nodes. Whereas all unexplored nodes are stored in the open list, the closed set stores all previously visited nodes. 
The open dictionary is used to improve the lookup of minimum $f$-values of already visited configurations. Initially, the open dictionary contains only the starting configuration (root node) of the storage bay. 

After initialization, the algorithm starts to search for a solution within the main loop and stops when an optimal solution without any blocking unit loads has been found or when the open list becomes empty, meaning no solution exists.
While nodes are in the open list and the time limit has not been reached, the node $n$ with the minimum cost estimation $f(n)$ is popped off the priority queue. If $n$ has already been visited and is in the closed set, then the node is skipped, as we already know the shortest path to get to it. Otherwise, the node is added to the closed set and the successor nodes for every possible move of a unit load are calculated, as long as they are not pruned due to branching rules (see \ref{branching}). Successor nodes are added to the priority queue and open dictionary if there is not already a node with the same configuration and a smaller cost in the closed set or open dictionary. This search procedure guarantees an optimal solution as long as the heuristic used in the cost estimation is an admissible lower bound on the true cost of the configuration.


\begin{algorithm}[!ht]
\SetAlgoLined
 \SetKwFunction{Fastar}{astar}
 \SetKwProg{Fn}{Function}{:}{}
 \Fn{\Fastar{root}}{
 open $\leftarrow$ PRIORITY\_QUEUE()\;
 open\_dict $\leftarrow$  $\emptyset$\;
 closed $\leftarrow$  $\emptyset$\;
 PUSH(open, root)\;
 open-dict(root) = $f($root$)$\;
 \While{open $>$ 0 and time $<$ timeout}{
  $n$ $\leftarrow$ POP(open)\;
  closed $\leftarrow$ closed\: $\bigcup$\: \{$n$\}\;
  \If{blocking($n$) $= 0$}{
   \KwRet $n$\; 
  }
 successors $\leftarrow$ BRANCHING($n$)\;
  \For{$s$ $\in$ successors}{
    \If{$s$ $\notin$ closed and ($s$ $\notin$ open\_dict or open\_dict($s$) $>$ f($s$))} {
     PUSH(open, $s$)\;
     open-dict($s$) = f($s$)\;
  }
  }
 }
 }
\caption{A* algorithm adapted from \cite{tierneySolvingPremarshallingProblem2017}}
\label{astar}
\end{algorithm}

\subsection{Branching} \label{branching}
For each iteration of the search, our algorithm calculates all feasible successors of the current node, i.e., configurations that can be reached by a single move of a unit load from the current configuration. Some of these configurations make no sense to examine, for example given node $n$, we should not reverse the move that brought us to node $n$. We further use \emph{memoization} to remember all seen configurations so that we do not explore the same configuration multiple times unless we find a cheaper way of reaching a configuration. 
\citet{tierneySolvingPremarshallingProblem2017} and \citet{tanakaSolvingRealworldSized2018} show that, in the context of CPMP, the use of domain-specific insights allows further improvements for branching, as there are usually many nodes that have the same $f$ value, thus one needs a way to differentiate between them.




\subsection{Lower bound on the number of moves}
In this section, we give an overview of the adapted lower bound from the CPMP and provide an example.
The notation 
is summarized in Table \ref{Tab:notation}. For mathematical proofs of correctness, we refer to the respective publications. 
\begin{table}[!ht]
\caption{Adopted notation from \cite{bortfeldtTreeSearchProcedure2012} \label{Tab:notation}}
\begin{tabularx}{\textwidth}{p{0.08\textwidth}X}
\toprule
  $n^{b}(w)$ & Number of blocking (badly placed) unit loads of the lane at column $i$\\
  $d(g)$ & Number of blocking unit loads represents demand of priority group $g$\\
  $D(g)$ & Cumulative demand of a priority group $g$ ($d(g) + d(g + 1) + \cdots + d(G)$)\\
  $s(g)$ & Number of supply slots for priority group $g$\\
  $S(g)$ & Cumulative supply of a priority group $g$ ($s(g) + s(g + 1) + \cdots + s(G) + H \cdot n_{l,\text{empty}}; n_{l, \text{empty}} – \text{number of empty lanes}$)\\
  $DS(g)$ & Cumulative demand surplus of priority group $g$ ($D(g)-S(g)$)\\
\bottomrule
\end{tabularx}
\end{table}
We adopt the classification of possible moves introduced by \citet{bortfeldtTreeSearchProcedure2012} into bad-good (BG), bad-bad (BB), good-good (GG) and good-bad (GB) moves. The notation indicates whether a unit load is blocking (bad) or well-placed (good) before and after a move. BX moves consist of BG and BB moves, whereas GX moves correspond to GG and GB moves.

\begin{figure}[!ht]
    \centering
    \includegraphics[width=13cm]{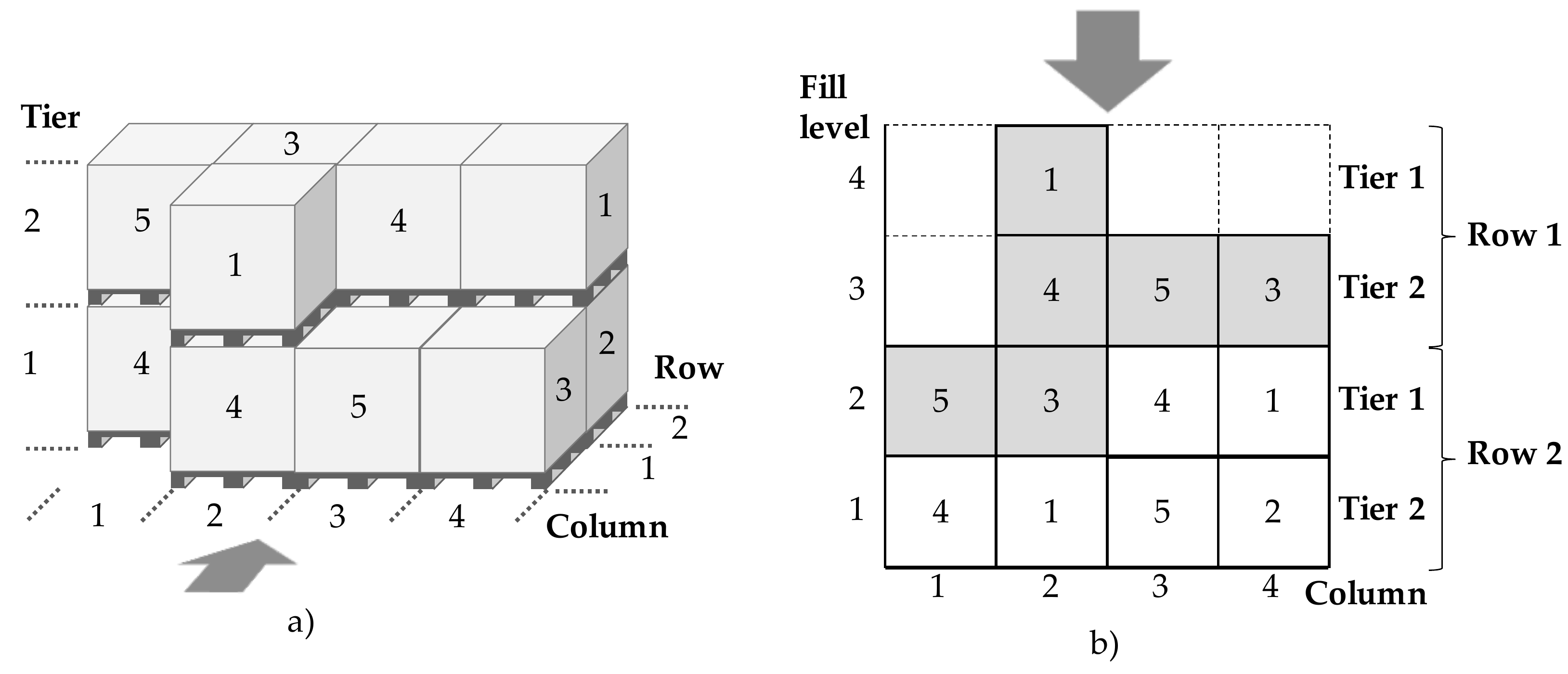}
    \caption{A simple example of a configuration for one access direction with a three-dimensional view on the left a), and translated into an access sequence for each lane at column $i$ on the right, b). Dark gray colored tiles represent blocking unit loads.}
    \label{fig:state_rep_example}
\end{figure}

In the case of a single access direction, the columns $I$ represent lanes, the rows $J$ refers to the depth level and $T$ is the number of tiers. Figure \ref{fig:state_rep_example} a) shows a small configuration with $I=4$, $J=2$ and $T=2$. Each lane at column $i$ can be translated into an access sequence of unit loads in Figure \ref{fig:state_rep_example} b). 
The tiles represent a storage position with the assigned group $g_{ijt} \in\{1,...,G\}$ of each unit load. All empty positions are set to $g_{ijt}=0$. All tiles of blocking unit loads that block other unit loads are colored in dark-gray.

For one access direction, a lane with a length of two stacks $J=2$ and two tiers $T=2$ results in a sequence of four storage locations that can be retrieved one after the other. We refer to the sequence of storage locations as the \textit{fill level}).
In the first lane at $i=1$ of Figure \ref{fig:state_rep_example} b), the unit load at position ($i=1$, $j=2$, $t=2$) is blocking, since it has a higher priority group than the unit load on the storage location below ($i=1$, $j=2$, $t=1$) with $g_{1,1,2}>g_{1,1,1}$. All unit loads that are placed at a storage location before a blocking unit load are also blocking, because they need to be moved for resolving all blockage. 
Hence, the unit loads at the positions ($i=2$, $j=1$, $t=2$), ($i=2$, $j=1$, $t=1$), ($i=2$, $j=2$, $t=2$), ($i=3$, $j=1$, $t=1$) and ($i=4$, $j=1$, $t=1$) are blocking as well.

\subsubsection{Simple lower bound:}
A first valid lower bound for the minimum number of item moves simply counts the number of blocking unit loads of the current bay configuration (see  \citet{LEE2009468, exposito-izquierdoPreMarshallingProblemHeuristic2012}). 
Since the unit load of higher priority will be retrieved first, at least all unit loads in front of it must be relocated. In the best case, all unit loads are relocated as BG moves from badly to well placed. In Figure \ref{fig:state_rep_example} b), the total count of blocking unit loads is $n_{b} = 6$. 

\subsubsection{Extension by \citet{bortfeldtTreeSearchProcedure2012}:} 
The lower bound for the number of moves from \citet{bortfeldtTreeSearchProcedure2012} is defined as the sum of the number of BX and GX moves. 
The first extension to the simple lower bound is that to carry out a BG move, it is necessary to have at least one lane without any badly placed unit loads. Otherwise, at least the lane with the smallest number of badly placed items has to be moved as a BB move $n_h = \min\{n_b{(s)}\}$. 
Therefore, the number of BX moves is the sum of the blocking unit loads (simple lower bound $n_{b}$) and the minimum number of blocking unit loads of all stacks $n_h$. 
In Figure \ref{fig:state_rep_example} b), all lanes contain badly placed items. Three lanes contain only one badly placed item. Therefore, $n_h = 1$ and a single additional move is necessary. 


The second extension refers to the number of GX moves, which involve moving well-placed unit loads. By calculating the required slots for badly placed unit loads (demand) as well as the available slots (potential supply) for each priority group, we derive a cumulative demand surplus. A slot is a storage location defined by the coordinates $w$, $l$ and $t$. The required demand $d(g)$ is computed by counting the number of badly placed items for each priority group. Table \ref{Tab:demand_surplus} shows $d(g)$ for the example in Figure \ref{fig:state_rep_example}.
The cumulative demand $D(g)$ is calculated from the lowest to highest priority groups. 

Once $D(g)$ has been determined, the supply slots $s(g)$ must be computed. The supply slots of a lane at column $i$ consist of the slots before the last well-placed item 
that do not need to be moved. The available slots for the priority group $g$ of the last well-placed item of each lane are added to the potential supply slots. For the example in Figure \ref{fig:state_rep_example} b) and Table  \ref{Tab:demand_surplus}, there 
are five potential supply slots, respectively, above $g=4$ (lanes 1 and 3) and $g=1$ (lanes 2 and 4). Again, the cumulative values $S(g)$ are calculated.

Finally, the resulting cumulative demand surplus is the difference of cumulative demand subtracted 
by the cumulative supply, since the unit loads with higher priority can always be stored on slots with lower priority. A positive cumulative demand surplus means that there are not enough available slots by just moving badly placed items. In this case, additional GX moves are necessary. The necessary GX moves are calculated by first deriving the maximum demand surplus for all priority groups $DS(g^*) = argmax\{DS(g)\}$. In our example in Table \ref{Tab:demand_surplus} the maximum demand surplus $DS(g^*)$ is $2$.

\begin{table}[]
\centering
\begin{tabular}{ M{1.75cm}M{1.75cm}M{1.75cm}M{1.75cm}M{1.75cm}M{1.75cm}  }
\hline
 Item group g & Demand d(g) & Cumulative demand D(g) & Potential supply s(g) & Cumulative potential supply S(g) & Cumulative demand surplus DS(g)\\
\hline
 5 & 2 & 2 & 0 & 0 & \textbf{2}\\
 4 & 1 & 3 & 5 & 5 & -2\\
 3 & 2 & 5 & 0 & 5 & 0\\
 2 & 0 & 5 & 0 & 5 & 0\\
 1 & 1 & 6 & 5 & 10 & -4\\
 \hline
\end{tabular}

\caption{\label{Tab:demand_surplus} Calculating demand surplus based on the example in figure \ref{fig:state_rep_example} b)}
\end{table}

We use the cumulative demand surplus $DS(g^*)$ to calculate the potential GX lanes $n^{pi}_{GX}$, where additional slots for unit loads of priority groups $g \geq g^*$ can be created. Slots for these priority groups can be created by relocating correctly placed unit loads of a priority group $g < g^*$. This means a potential GX lane must contain correctly placed unit loads of a smaller priority group. 
The correctly placed unit load of the smallest priority group of each lane corresponds to the outermost correctly placed item.
Empty lanes are not potential GX lanes, since their supply is already fully taken into account with the highest priority group and no additional slots can be created.
The necessary number of GX lanes must be at least 
$n^i_{GX} = max(0,[DS(g^*)/(J \cdot T)])$ with a maximum of $J \cdot T$ available slots per lane, if a whole lane has to be cleared. This is a slight modification compared to the approach for the CPMP with only $T$ slots per stack. 
The number of potential GX lanes $n^{pi}_{GX}$ must be greater or equal than the necessary number of GX lanes $n^i_{GX}$ to find a feasible solution. If this is the case, the lanes are sorted by  $n^{g^*}(i)$ in ascending order. 
$n^{g^*}(i)$ is the number of correctly placed unit loads with priority group $g < g^*$ of a lane at column $i$. In our example of Figure \ref{fig:state_rep_example} b), lanes 1, 2 and 3 contain one correctly placed item and lane 4 contains two correctly placed items smaller than $g^*=5$. This leads to the number of potential GX lanes $n^{pi}_{GX} = 4$ and we get the sorted list of correctly placed items that have to be moved $[1, 1, 1, 2]$.  
The lower bound for the number of GX moves is calculated by $n_{GX} = \sum^{n^i_{GX}}_{j=1}n^{g^*}(i)$. 
For the example in figure \ref{fig:state_rep_example} b), we round the result of $max(0, [2/(2*2)]$ up to $n^i_{GX} = 1$ and derive $n_{GX} =  \sum^{1}_{j=1}n^{g^*}(i) = 1$. 

Further improvements for cases in which the lower bound can be increased have been proposed in \citet{tanakaSolvingRealworldSized2018} and \citet{tanakaBranchBoundApproach2019}. These improvements can be transferred analogously, but due to their complexity we omit them from this study. 

\section{Novel approach for solving the UPMP with multiple access directions} 
\label{5_UPMP_two}

In storage bays with multiple access directions, each stack can generally be accessed from any allowed access direction. 
Multiple access directions are beneficial and challenging at the same time. 
An advantage of multiple access directions is that potentially fewer relocation moves are required to sort the storage bay until each unit load can be accessed from one access direction without blockage. However, difficulties are a tremendous increase in the branching factor as well as 
the solution space and a higher complexity for configuration assessment. 

To overcome these challenges, we propose a solution approach in which we initially determine the access direction for each stack and fix this direction until the pre-marshalling process is finished (access direction fixing). This allows us to benefit from a significantly reduced number of required moves and, at the same time, achieve fast runtimes. 
Access direction fixing makes also sense from a practical point of view. We will discuss this together with the implications of our assumption 
at the end of the section.

\begin{figure}[!ht]
    \centering
    \includegraphics[width=7cm]{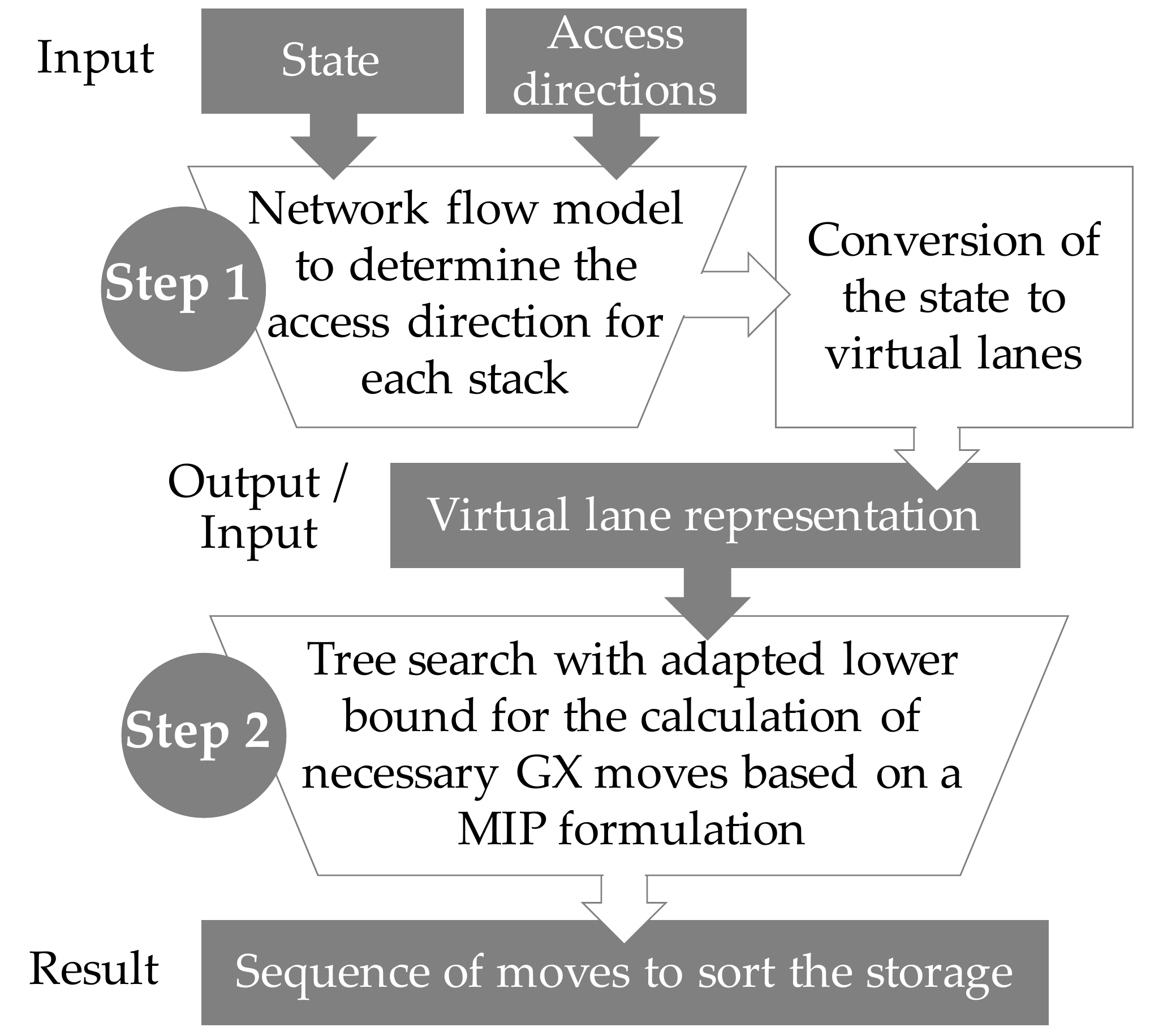}
    \caption{Illustration of the main components of our solution approach with the respective input and output.}
    \label{fig:novel_solution_approach}
\end{figure}

For solving the UPMP with multiple access directions, we present a novel two-stage algorithm shown in Figure  \ref{fig:novel_solution_approach}. 
First, a network flow model is used to initially determine the access direction fixing for each stack that minimizes the number of blocking items. This configuration is transformed into a representation of \textit{virtual lanes} similar to the single access direction case and is used throughout the search. A virtual lane consists of 1 to $\max (I,J)$ successive stacks in a row/column with a defined access direction. 
Since the length of each virtual lane can be different, we introduce an adapted lower bound with a mixed-integer programming (MIP) formulation to determine the minimum number of required reshuffling moves.

\begin{figure}[!ht]
    \centering
    \includegraphics[width=12.5cm]{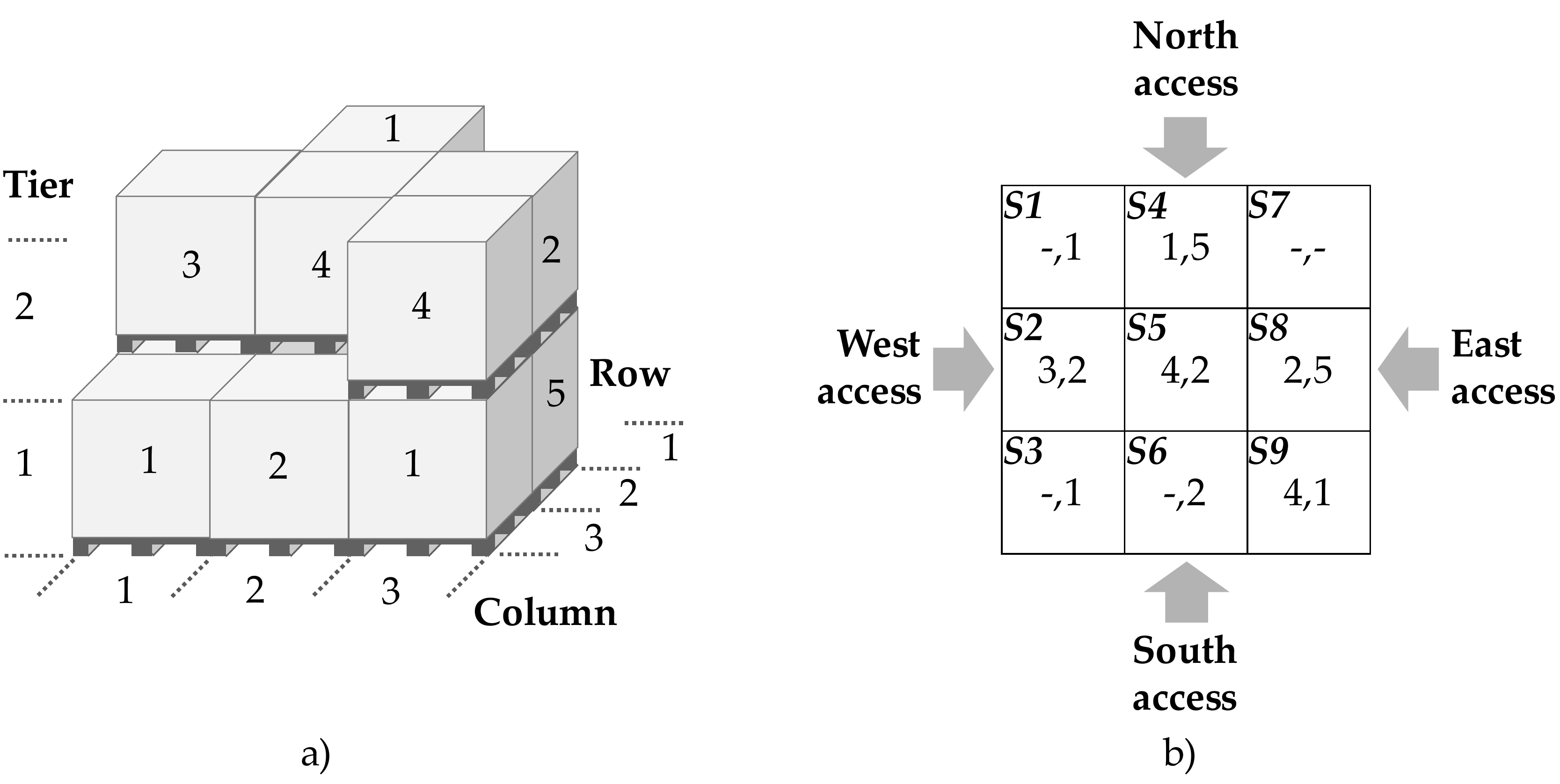}
    \caption{Example of a bay configuration with four access directions in a). Translated into a top view in b), each tile in the grid represents a stack with a tuple of multiple tiers. Hyphens 
    indicate empty storage locations.}
    \label{fig:four_directions_example_grid}
\end{figure}



\textbf{Step 1: Access Direction Fixing}\\ 
We first describe the components of the mathematical model for calculating the access direction for each stack and then provide an example based on Figure \ref{fig:four_directions_example_grid}. Our model captures the grid-based layout of a block stacking storage system and considers simple rules of feasible AMR movements. Table \ref{tab:notation} shows the sets, parameters, and decision variables that we use throughout this section.  
Figure \ref{fig:four_directions_example_grid} a) shows a simple example of the configuration with $I=3$ columns, $J=3$ rows and two tiers $T=2$. 
Figure \ref{fig:four_directions_example_grid} b) translates the example into the associated top view. Each tile of the grid represents a stack $S1$ to $S9$ with a tuple of stacked unit loads (tier $2$, tier $1$). 

\begin{table}[!ht]
\caption{Notation of the network flow model}
\begin{tabularx}{\textwidth}{p{0.3\textwidth}p{0.6\textwidth}}
\toprule
  \multicolumn{2}{l}{{Sets and Indices}}\\
  \midrule
  $s,t$ & Indices of network vertices\\
  $d$ & Index of the neighbor vertex for a cardinal access direction ($\{north, south, east, west\}$)\\
  $n$ & Index of an orthogonal neighbor vertex of $x_{d,s}$\\
  $E$ & Set of all edges\\
  $E^{x}$ & Set of edges excluding edges from the origin $o$\\
  $V^{x}$ & Set of vertices excluding the origin $o$\\
  $E^\mathit{origin}$ & Set of edges from origin $o$\\
  $V^\mathit{origin}$ & Set of vertices of origin $o$\\ 
  $V^\mathit{inner}$ & Set of inner vertices\\
  \midrule
  \multicolumn{2}{l}{{Parameters}}\\
  \midrule
  $c_{s,t}$ & Cost of visiting vertex $t$ from vertex $s$\\ 
  $b_{s} \in \mathbb{N}$ & Demand at vertex $s$ ($1$ for each stack that has to be visited)\\ 
  $S \in \mathbb{N}$ & Supply provided at origin $o$ (equal to the number of stacks)\\
  \midrule
  \multicolumn{2}{l}{{Variables}}\\
  \midrule
    $f_{s,t}$, $f_{t,s}$ & Flow from source $s$ to destination $t$ and vice versa\\
    $x_{s,t} \in\{0,1\}$ & Equals 1, if a stack is accessed via source $s$ to destination $t$\\ 
    $x_{d,s} \in\{0,1\}$ & Equals 1, if stack $s$ is accessed via the edge of access direction $d$\\ 
    $x_{s,n} \in\{0,1\}$ & Equals 1, if the outgoing edge from stacks $s$ to an orthogonal neighbor $n$ is used\\ 
\bottomrule
\end{tabularx}
\label{tab:notation}
\end{table}

In our model, each stack is represented by a vertex with a directed edge from and to the adjacent stack vertices for any allowed access direction. For each of the allowed access directions (up to four) a vertex is added with edges pointing to the vertices of stacks that are placed at the edge of the bay.  
Finally, a vertex representing the origin $o$ is connected to vertices for each access direction.  
The capacity of edges is unlimited and the assigned costs represent the incremental difference of blocking items between accessing the previous and the current stack in a lane. To avoid holes, i.e., items in front of empty locations, we add a cost of one for any hole that may be created. Holes are detrimental as they lead to less free storage locations in front of the items and, therefore, to infeasible configurations. 

Supply is provided at origin $o$ for all of the stacks' vertices. Each vertex representing a stack 
must be visited once, and thus has a demand of $1$. The intermediate vertices for the access directions have zero demand. Edges between vertices that are directly accessible by an access direction are not necessary, since the minimum costs in terms of blocking items are guaranteed through access from the respective direction. The optimization objective is to minimize the costs of accessing all stack vertices (demand) from supply vertex $o$ via intermediate vertices for the available access directions. The result is a defined access configuration with a minimum number of blocking unit loads. 
The mathematical formulation of our network flow model is:
\begin{gather}\label{eq:1} 
min \sum_{(s,t) \in E} c_{s,t} x_{s,t} 
\end{gather}
\begin{gather}\label{eq:2} 
f_{s,t} \leq P x_{s,t}\quad \forall (s,t) \in E
\end{gather}
\begin{gather}\label{eq:3} 
\sum_{(s,t) \in E^x} f_{s,t} - \sum_{(t,s) \in E^x} f_{t,s} = b_s\quad \forall s \in V^x
\end{gather}
\begin{gather}\label{eq:4} 
\sum_{(s,t) \in E^\mathit{origin}} f_{s,t} = P
\end{gather}
\begin{gather}\label{eq:5} 
\sum_{(d,s) \in E} x_{d,s} + \sum_{(s,n) \in E} x_{s,n} \leq 1 \quad \forall s \in V^\mathit{inner}
\end{gather}
\begin{gather}\label{eq:6} 
x_{s,t} \in \{ 0,1\},\quad f_{s,t} \geq 0\quad \forall (s,t) \in E
\end{gather}


The objective in term~\eqref{eq:1} is to minimize the movement costs for accessing all stacks.  
We use the binary variable $x_{s,t}$ to indicate whether an arc is used or not. 
The idea is that we want to count the number of blocking items only once, even if the stacks behind are accessed through the same stack. Constraints~\eqref{eq:2} ensures no arc flow exceeds the total amount of supply $S$. 
Constraints~\eqref{eq:3} and~\eqref{eq:4} ensure flow conservation for all vertices with a demand of zero and above, as well as flow conservation for supply in the origin $o$. 
We require Constraints~\eqref{eq:5} to avoid the AMRs turning within the grid.  The constraints say that flow on orthogonal neighbor edges is not allowed for all inner vertices. 
Inner stacks are all stacks that cannot be directly accessed by one of the access direction vertices. We note that the number of constraints to avoid turning increases with the grid size (up to eight per inner stack). If the AMRs are able to move around corners in the grid, these constraints can be neglected. Our mathematical model is similar to a min-cost flow formulation, 
however, we use binary variables instead of flow variables in our objective function because the costs should only be considered once and add additional constraints in~\eqref{eq:5} for AMR movements.

\begin{figure}[!ht]
    \centering
    \includegraphics[width=14cm]{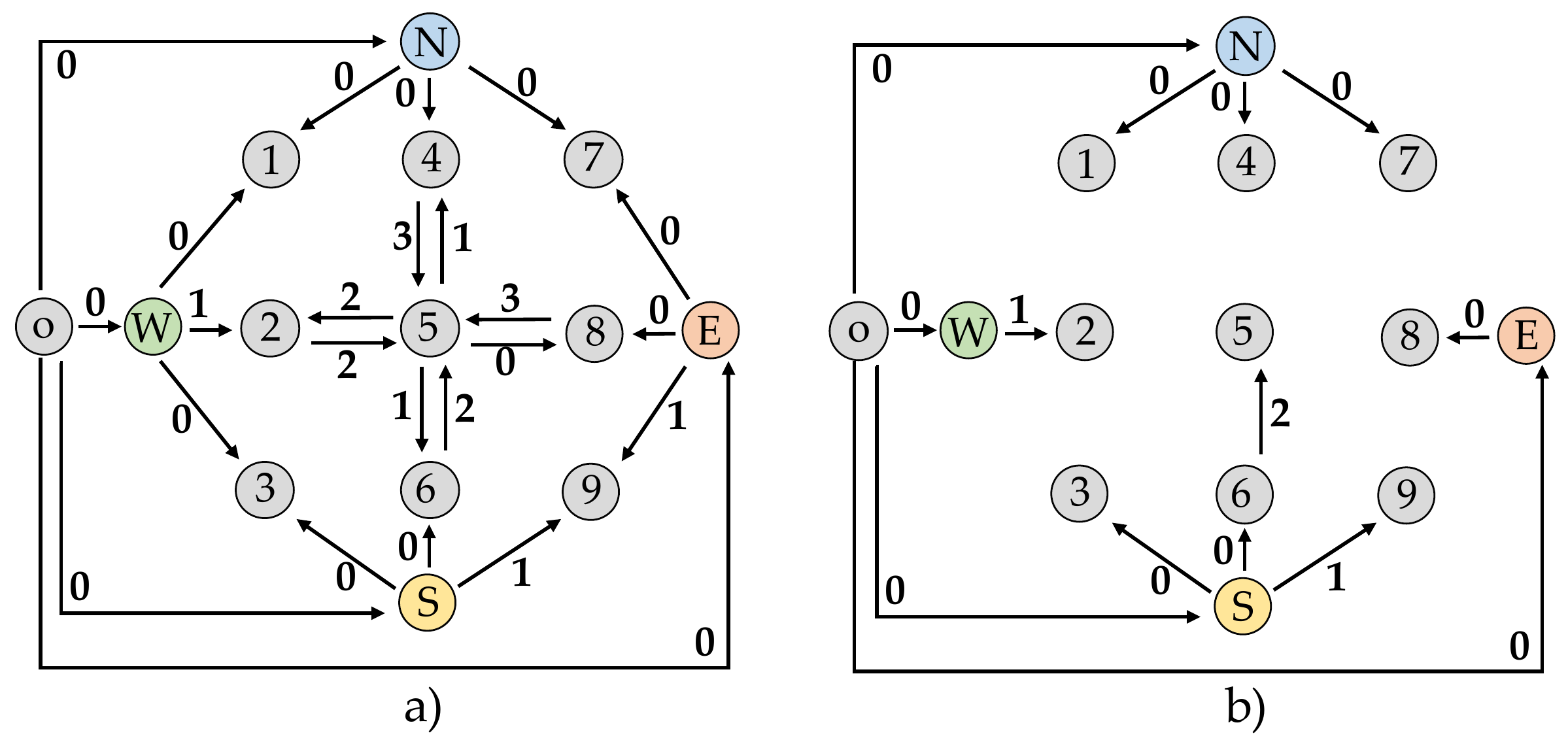}
    \caption{Network flow scheme for the example with four access directions of Figure \ref{fig:four_directions_example_grid} in a) and a possible solution with fours blocking items in b). Each arc is labeled with its cost.} 
    \label{fig:Example_network_flow}
\end{figure}

\paragraph{Example} We now consider the example given in Figure \ref{fig:four_directions_example_grid} and derive a representation for the network flow model. 
All stacks represented by tiles are translated into vertices S1 to S9 in the network flow representation in Figure~\ref{fig:Example_network_flow} a). For each access direction a vertex $N$, $S$, $W$ and $E$ is added, as well as a vertex for the origin $o$. Edges between the outermost stacks of the bay for any allowed access direction are not necessary. These stacks can be directly accessed from the respective direction. 
Each edge is assigned costs that are equal to the incremental cost of of accessing the corresponding vertex.
In our example, accessing stack $S2$ from $W$ costs one due to one blocking item. Now accessing stack $S5$ from stack $S2$ would add another two blocking items, as both items of stack two would have to be moved plus the top item of stack $S5$. 
Supply is provided at origin $o$ for all stacks (in our example: 9 vertices). 
Each stack must be visited once and therefore has a demand of $1$. The intermediate vertices for the access directions have zero demand. 
Figure \ref{fig:Example_network_flow} b) illustrates a solution for a lower bound of blocking unit loads showing only used edges.

\begin{figure}[!ht]
    \centering
    \includegraphics[width=\textwidth]{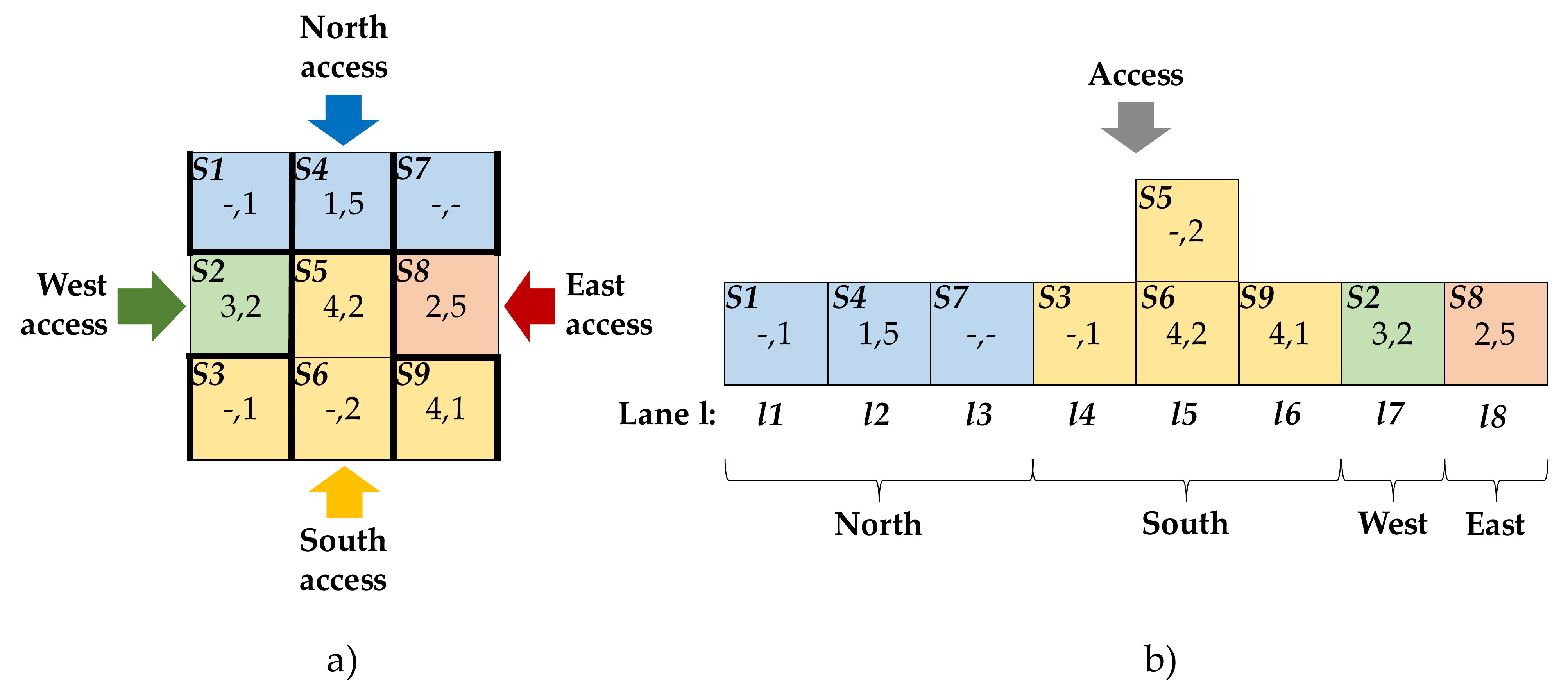}
    \caption{Example from Figure  \ref{fig:four_directions_example_grid}. Thick black lines illustrate the virtual lanes in a). In b) these lanes are transformed into a representation of one access direction.}
    \label{fig:state_transformation_to_lanes}
\end{figure}

After the access direction for each stack is determined, the configuration is represented as virtual lanes and fixed for the search procedure. Figure \ref{fig:state_transformation_to_lanes} a) shows the grid representation with thick black lines to illustrate the virtual lanes. In Figure \ref{fig:state_transformation_to_lanes} b) these lanes are transformed into a representation of one access direction. The corresponding access directions are specified underneath the stacks.\newpage

\textbf{Step 2 - Tree search in virtual lanes}\\ 
For the lower bound heuristics, we build on the adapted lower bound for one access direction: 

We use the same procedure as for one direction to calculate the BX moves as well as to derive 
the demand, supply and the demand surplus $DS(g^*)$ and $g^*$. 
After the demand surplus $DS(g^*)$ has been calculated, the next step is to determine the necessary number of GX lanes $n_{GX}$. This cannot be calculated in the UPMP via $n^i_{GX} = max(0,[DS(g^*)/(J \cdot T)])$, since the length of each virtual lane can be different. 
Also calculating the number of GX moves via $n_{GX} = \sum^{n^i_{GX}}_{j=1}n^{g^*}(i)$ is problematic. It could make sense to select a lane that is more costly in terms of GX moves than another lane, but at the same time frees up more space. 
In fact, we need to select the lanes where a minimum number of GX moves is necessary to free up enough space for $DS(g^*)$. Therefore, we use a MIP formulation that minimizes the number of moves. The notation is shown in Table \ref{tab:notation_MIP}.

\begin{table}[!ht]
\caption{Notation of the MIP model}
\begin{tabularx}{\textwidth}{p{0.3\textwidth}p{0.6\textwidth}}
\toprule
  \multicolumn{2}{l}{{Sets and Indices}}\\
  \midrule
  $L$ & Number of virtual lanes ($l \in\{1,...,L\}$) with well-placed items of a priority group $g<g^*$)\\
  \midrule
  \multicolumn{2}{l}{{Parameters}}\\
  \midrule
  $k_{l}$ & Cost of moving all well-placed items with a priority group $g<g^*$\\
  $p_{l}$ & Potentially freed space of lane $l$ calculated via the length of each lane $l$ minus the well-placed items with $g>=g^*$\\ 
  \midrule
  \multicolumn{2}{l}{{Decision variables}}\\
  \midrule
  $y_{l} \in\{0,1\}$ & Binary decision variable, if virtual lane $l$ is selected to be emptied\\ 
\bottomrule
\end{tabularx}
\label{tab:notation_MIP}
\end{table}

The MIP model to calculate GX moves is:
\begin{gather}\label{eq:7} 
min \sum_{l \in L} k_{l} y_{l}
\end{gather}
\begin{gather}\label{eq:8} 
\sum_{l \in L} y_{l} p_{l} \geq DS(g^*)
\end{gather}
\begin{gather}\label{eq:9} 
y_{l} \in \{ 0,1\}\quad \forall l \in L
\end{gather}

The objective function in term~\eqref{eq:7} minimizes the sum of costs of the selected lanes.  
Constraint~\eqref{eq:8} ensures that the space requirement is fulfilled, which means that the sum of freed up storage slots must be equal or higher than the demand surplus $DS(g^*)$. 

Applying the proposed lower bound on the example of Figure \ref{fig:state_transformation_to_lanes}, we get four BX moves. For the demand surplus, $DS(g^*) = 1$ and $g^* = 3$. Next, we calculate the costs $k_l$ to empty each lane, 
as well as the potentially freed up space $p_l$ of each lane $l$. Table \ref{Tab:example_multiple_directions} shows the results of $k_l$ and $y_l$ for each lane of our example. The lanes $l$ with no well-placed items of a priority group $g<g^*$ are not considered because no additional space for $g^*$ can be created. Therefore, lane $l3$ can be neglected.

\begin{table}[]
\centering
\begin{tabular}{ M{1.00cm}M{1.00cm}M{1.00cm}M{1.00cm}M{1.00cm}M{1.00cm}M{1.00cm}M{1.00cm}M{1.00cm} }
 \hline
 Parameter type & $l1$ & $l2$ & $l4$ & $l5$ & $l6$ & $l7$ & $l8$\\
 \hline
 $k_l$ & 1 & 1 & 1 & 1 & 1 & 1 & 1\\ 
 $p_l$ & 2 & 1 & 2 & 4 & 2 & 2 & 1\\
 \hline
\end{tabular}
\caption{\label{Tab:example_multiple_directions} $k_l$ and $p_l$ for each lane $l$ of the example.}
\end{table}

In this simple example, any lane besides $l3$ would free up the required one storage location for items with $g>=g^*$. Since the costs are for all lanes the same, one of them can be chosen at random. We add one GX move to the lower bound and get an estimation of overall five moves to sort the storage. 

\subsection{Impact of access direction fixing}
\label{sec:impact_access_direction_fixing}
\begin{figure}[!ht]
    \centering
    \includegraphics[width=10.00cm]{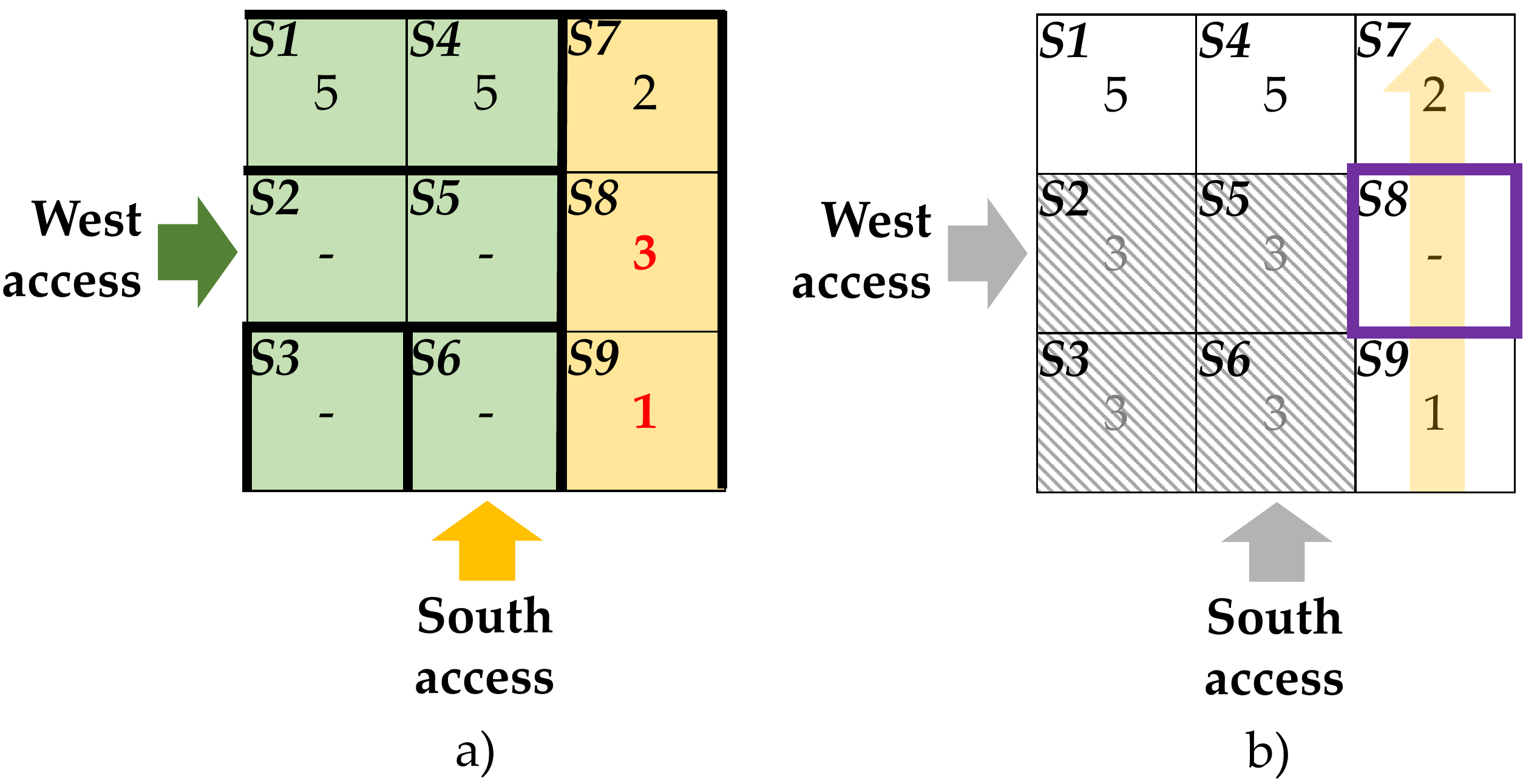}
    \caption{The example in a) shows a configuration with access direction fixing
    illustrating two blocking unit loads in bold red. In b) all blockage is resolved by moving the unit load of stack $S8$ (box with thick purple lines) to a possible location indicated with crosshatched tiles and a light gray number.}
    \label{fig:problem1}
\end{figure}

\begin{figure}[!ht]
    \centering
    \includegraphics[width=\textwidth]{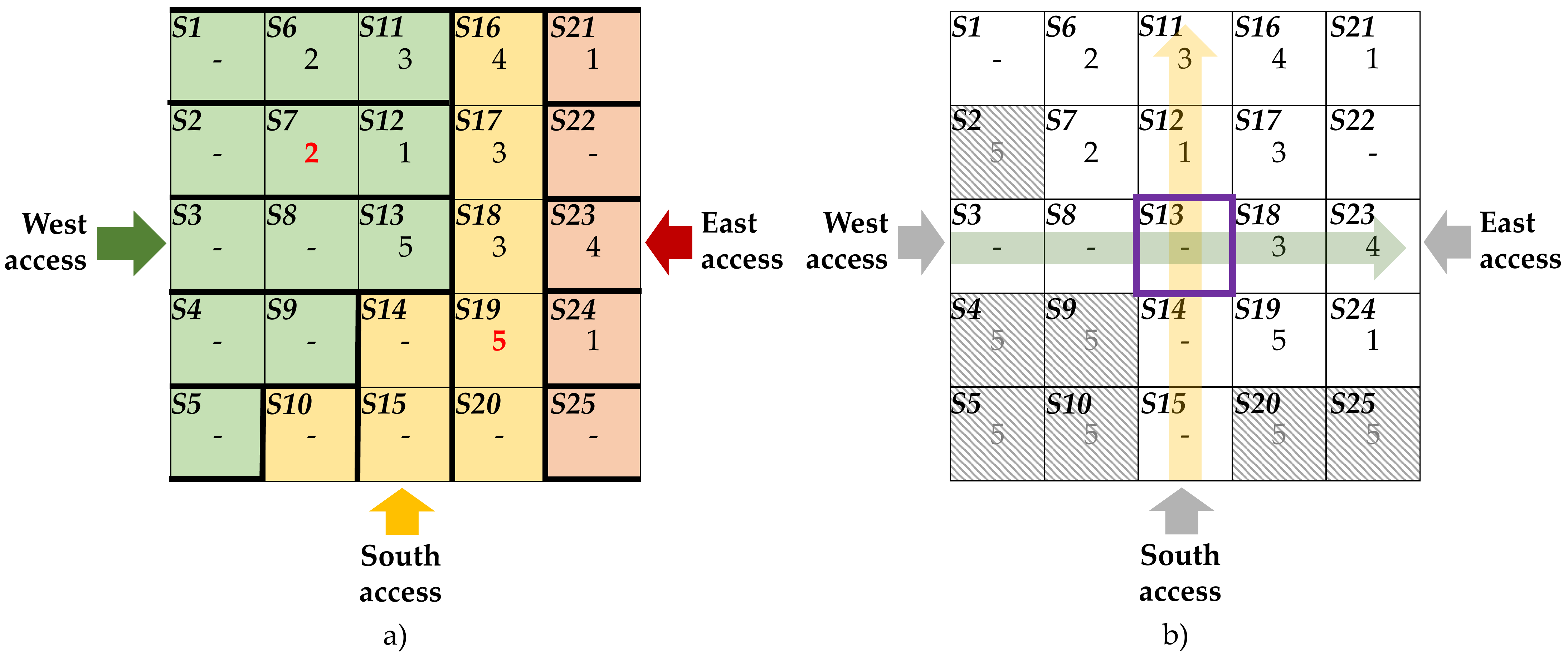}
    \caption{Example in a) shows a configuration with access direction fixing 
    illustrating two blocking unit loads in bold red. In b) all blockage is resolved by moving the unit load of stack $S13$ (box with thick purple lines) to a possible location indicated with a crosshatched tile and a light gray number.}
    \label{fig:problem2}
\end{figure}

As indicated at the beginning of the section, in our problem definition we fix the access directions for each stack over the course of the search. Restricting the solution space might increase the required number of moves in certain situations.
We show this effect based on two examples in Figure \ref{fig:problem1} and \ref{fig:problem2}. 

The example in Figure \ref{fig:problem1} a) shows the result of the network flow model for corner access from south and west. The unit load of priority group $2$ at stack $S7$ is blocked by two unit loads in the stacks $S8$ and $S9$. However, this blockage could be resolved by a single move if the access directions are not fixed. 
Figure \ref{fig:problem1} b) shows possible places to relocate the load in stack $S8$ (light-gray crosshatch) when accessing it from the west. 
This allows access to the load in stack $S7$ from the south without any blockage by passing over the empty stack $S8$. Thus, stack $S8$ is accessed from two different access directions over the course of the pre-marshalling operation. Placing an item at $S2$ or $S5$ is, however, problematic, as this leads to a hole as the empty stack $S8$ cannot be directly accessed anymore. Additional rules to prevent holes are necessary, 
but this limits the use of available storage locations.

Figure \ref{fig:problem2} a) is another example with three access directions, where the items at stack $S12$ and at stack $S18$ are blocked. The network flow model would propose to relocate two blocking items in front (e.g., $S7$ and $S19$), but again, as shown in \ref{fig:problem2} b), a solution is possible via a single move in case there is no access direction fixing. The item from stack $S13$ can be moved via west or south access to the light-gray crosshatched locations. Stacks $S12$ and $S18$ can be accessed via stack $S13$ from the south and west. The difficulty is to maintain this access over time: 
When placing a unit load at stack $S13$, it must be ensured that it does not block any other unit loads in both access directions.

These two examples show that fixing the access directions may require additional moves compared to solutions where a stack can be visited from several access directions over time.
This means that the network flow model is not a valid lower bound without access direction fixing. However, real-life use-cases benefit from access direction fixing 
due to easier traffic management and a more stable and coherent configuration, especially for changes of priorities and incoming unit loads. 
To assess the share of instances, where additional moves are required compared to the network flow model further investigation is required.

\section{Computational experiments}
\label{6_experiments}
The goal of our analysis is to evaluate the effects of different access directions and assess the performance of the proposed solution approaches for the UPMP. Therefore, we introduce a publicly available dataset at \url{https://doi.org/10.5281/zenodo.6791321} and present the results in terms of the number of solved instances, runtime, and total moves. We further analyze the the gap of our lower bound at the root node, as well as provide insights into the value of multiple access directions.
We conduct our computational experiments on AMD EPYC 7401P CPUs at 2.45 GHz. 
Our algorithms are implemented in Python applying Gurobi 9.5 with the default parameters for both the network flow model and the MIP. The execution time for each instance is limited to one hour of CPU time. 


\subsection{Instance generator and dataset} 
We develop an instance generator for the UPMP that places unit loads in a storage bay depending on the defined access directions. In addition to allowing up to four access directions (north, south, east and west), we vary the number of columns, rows and tiers of the storage bay, as well as the fill percentage. The instance sizes are inspired by real-world block stacking storage systems in production and logistics. 
We model five priority groups that can occur with uniform probability. We place unit loads 
in the storage bay successively at random and generate a new instance 
for each possible combination. During this procedure, it is necessary to consider the defined access directions to provide free storage locations that are directly accessible and avoid gaps. 
The generated instances may be infeasible, especially for high fill percentages. 
%
We generate a dataset with 1650 instances in which there are ten instances for each possible combination of the parameters shown in Table \ref{Tab:dataset_parameter}.  

The access variants (combinations of possible access directions) are: 
\begin{itemize}
    \item north (referred to as \textit{single})
    \item north and west (referred to as \textit{corner}) 
    \item north and south (referred to as \textit{opposite})
    \item north, south and west (referred to as \textit{three})
    \item all directions (referred to as \textit{four})
\end{itemize}

\begin{table}[]
\centering
\begin{tabular}{ M{3.50cm}M{6.50cm}}
 \hline
 Parameter & Value\\ 
 \hline
 Access directions & \{single, corner, opposite, three, four\}\\ 
 Size (L\,x\,W\,x\,T) & \{3x3x1, 4x4x1, 5x5x1, 6x6x1, 7x7x1, 8x8x1, 9x9x1, 10x10x1, 3x3x2, 4x4x2 5x5x2\}\\ 
 Fill percentage & \{40\%, 60\%, 80\%\}\\
 \hline
\end{tabular}
\caption{\label{Tab:dataset_parameter} Varied parameters of the generated benchmark dataset}
\end{table}

Besides the sizes with a quadratic footprint shown in Table \ref{Tab:dataset_parameter} other rectangular footprints are possible but omitted from this study due to limited space. 
The dimensions start at a 3x3x1 footprint (L\,x\,W\,x\,T) up to a size where we reach the limit of solvability due to timeouts for all access variants. A maximum 10x10x1 footprint for a single tier (overall 1200 instances: 10 instances for 5 access variants, 8 sizes and 3 fill levels) and a 5x5x2 footprint for two tiers (overall 450 instances: 10 instances for 5 access variants, 3 sizes and 3 fill levels) results in up to 100 or respectively 50 storage locations for a single storage bay. 
This covers a large amount of real scenarios of small storage systems with a single bay.


\subsection{Runtime and number of instances solved}
Table~\ref{Tab:tier1_results_summary_results} provides an aggregated summary of the number of instances solved over all sizes and fill levels for a single tier. The detailed results are provided in Appendix~\ref{appendix_detailed_results}. 
The access variants are ranked in ascending order according to the number of solved instances, which represent all feasible solutions. 
In terms of infeasible instances, we are only able to verify infeasibility for instances of a single access direction and very small sizes 
when the fill level is at 80\%. 
We anticipate that there are some infeasible instances for other access variants as well, but we are unable to prove this within the timeout. 
Timeouts occur when we reach the limits of solvability for each access direction.

\begin{table}[]
\centering
\begin{tabular}{M{1.50cm}M{1.50cm}M{1.50cm}M{1.50cm}M{1.50cm}M{1.50cm}M{1.50cm}}
\toprule
Access variant &  No. solved &  No. infeasible &  No. timeout \\ 
\midrule
    single &       162 &           8 &       70 \\ 
    corner &       202 &           0 &       38 \\ 
    opposite &       219 &           0 &       21 \\ 
    three &       232 &           0 &        8 \\ 
    four &       237 &           0 &        3 \\ 
\bottomrule
\end{tabular}
\caption{\label{Tab:tier1_results_summary_results} Results summary for a single tier, showing the number of instances solved, proved infeasible and unsolved (in total 240 instances: 10 instances for 8 sizes and 3 fill levels).}
\end{table}

\begin{figure}[!ht]
    \centering
    \includegraphics[width=\textwidth]{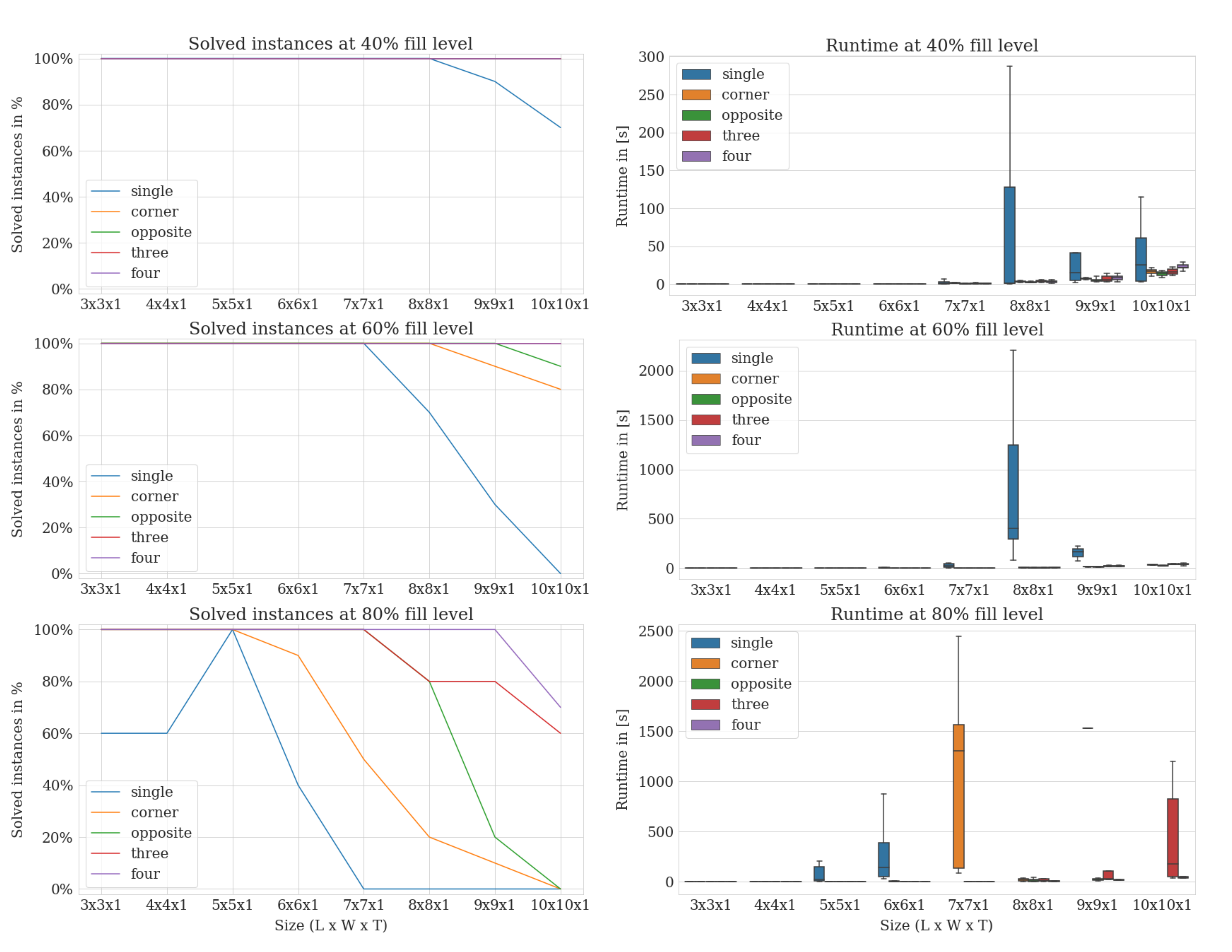}
    \caption{Overview of solved instances for a single tier and their runtimes for each fill level over the bay sizes calculated by L\,x\,W\,x\,T.}
    \label{fig:instances_runtime}
\end{figure}

Figure \ref{fig:instances_runtime} shows lineplots with the amount of up to ten solved instances (left column) and boxplots (without outliers) illustrating the runtime in seconds (right column) for each fill level (rows). Each plot provides the results for all access variants (legend) for each size of the dataset (x-axis).

The amount of solved instances goes down with increasing size and fill levels. With only a single access direction we can not solve all instances at a fill level of 40\% (infeasible and timeout), however with four access directions, we solve all instances up to a fill level of 80\% and size 10x10x1. Compared to single access, the runtimes for the access variants with multiple access directions stay at a very low level. The overall mean runtime decreases from 71.42 seconds for single access, to 56.07 seconds for corner, to 31.31 seconds for three, to 8.31 seconds for opposite and 7.72 seconds for four access directions. Note that the number of solved instances, including larger sizes, increases for multiple access directions. The maximum runtime for four access directions up to a size of 9x9x1 is 30.43 seconds. Even for 10x10x1 where not all instances could be solved anymore, we do not exceed 187.16 seconds.



At the 80\% fill level, finding a solution becomes more difficult: more infeasible configurations appear due to a limited amount of empty storage locations and, on average, more required moves are necessary to sort the block storage. For a single access direction we encounter four infeasible instances of sizes 3x3x1 and 4x4x1 (see also Table \ref{Tab:tier1_results_all}). The size 5x5x1 is the only size where all instances for each access direction can be solved. The reason for this is that 
the 80\% fill level leaves 5 empty locations for premarshalling, which is sufficient to clear a full lane. 
At the size 10x10x1, using three access directions results in six instances solved, and using four access directions does not allow for solving all instances. This overview shows in particular that enabling multiple access directions 
allows many more instances to be solved than using a single access direction. Furthermore, the higher numbers of directions lead to comparatively low runtimes. 

\begin{table}[]
\centering
\begin{tabular}{M{1.50cm}M{1.50cm}M{1.50cm}M{1.50cm}M{1.50cm}M{1.50cm}M{1.50cm}}
\toprule
Access variant &  No. solved &  No. infeasible &  No. timeout \\ 
\midrule
    single &        55 &          10 &       25 \\ 
    corner &        67 &           0 &       23 \\ 
    opposite &        82 &           0 &        8 \\ 
    three &        84 &           0 &        6 \\ 
    four &        87 &           0 &        3 \\ 
\bottomrule
\end{tabular}
\caption{\label{Tab:tier2_results_summary_results} Results summary for two tiers showing the number of solved, infeasible and unsolved instances (in total 90 instances: 10 instances for 3 sizes and 3 fill levels)}
\end{table}

Table~\ref{Tab:tier2_results_summary_results} presents a short summary of the results for two tiers. Detailed results are again provided in Appendix~\ref{appendix_detailed_results}. Our experiments show that adding additional tiers in a scenario with a fully shared storage strategy is very demanding. Already at a size limit of 5x5x2, we are not able to solve all instances with any access variant. Nevertheless, multiple access directions improve the situation: The number of solved instances increases from 55 for a single access direction to 87 out of 90 for four access directions. 

\begin{figure}[!ht]
    \centering
    \includegraphics[width=0.6\textwidth]{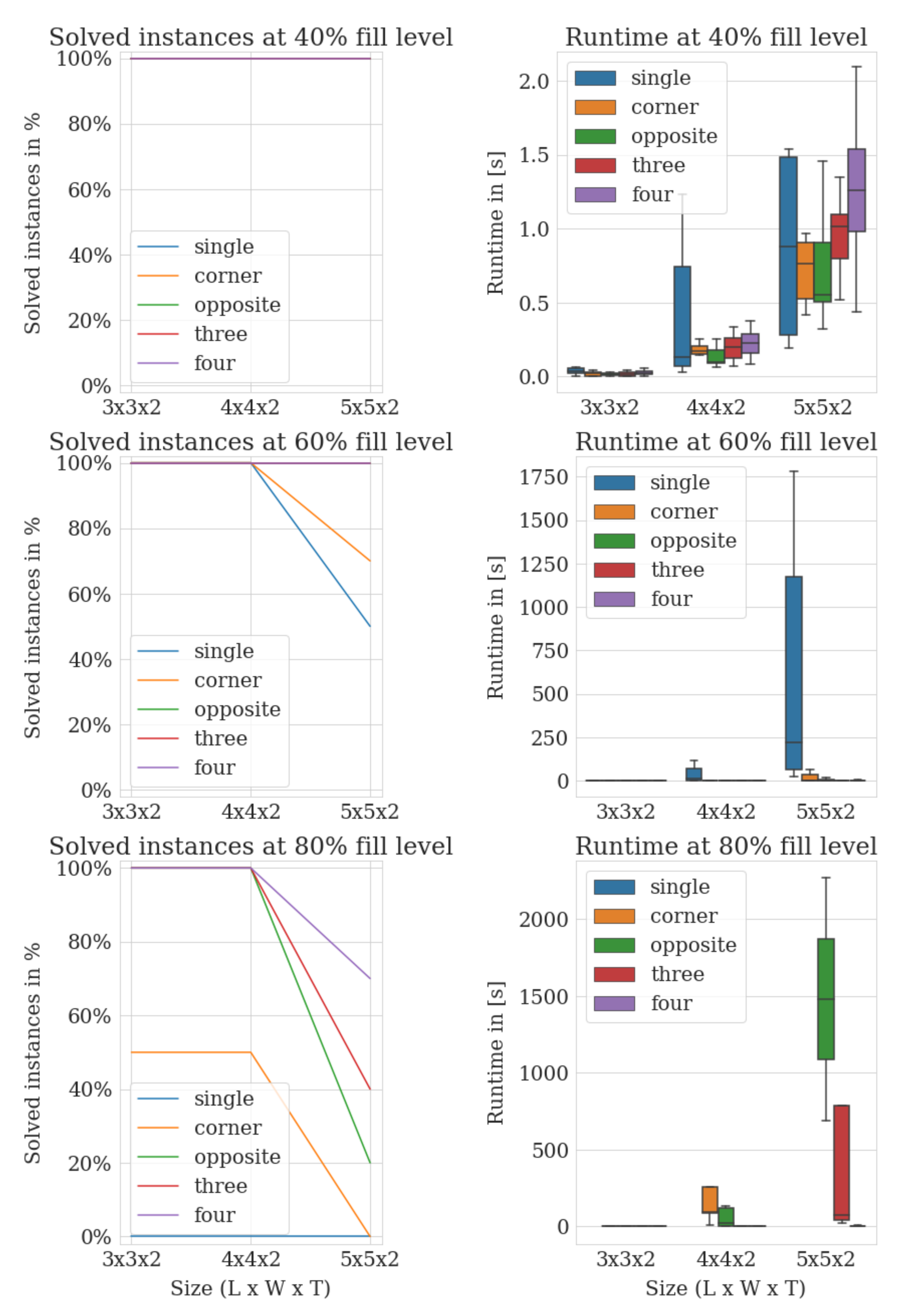}
    \caption{Overview of solved instances with two tiers and their runtimes for each fill level over the bay sizes calculated by L\,x\,W\,x\,T.}
    \label{fig:instances_runtime_two_tiers}
\end{figure}

Figure \ref{fig:instances_runtime_two_tiers} shows the plots for solved instances and their runtime for two tiers. For a fill level of 40\%, all instances are solved with overall very low runtimes. The runtimes are even slightly increasing with the amount of access directions due to an increased branching factor (more virtual lanes). For higher fill levels, the runtime is decreasing with an increasing amount of access directions due to less moves that are required. 
For two tiers, we are able to solve all instances up to size 4x4 and 80\% fill level with two opposite, three and four access directions. Finding a solution for corner access is already difficult for very small instances, since an unsorted stack in the back corner may lead to an infeasible instance that cannot be solved if not enough empty storage slots are available. Even though we are only able to solve rather small instances, the improvement of multiple to a single access direction is evident as a single direction fails to solve any instance with 80\% fill level. The overall mean runtimes improve from 101.19 seconds for single access, to 93.82 seconds for corner, to 60.80 seconds for three, to 57.03 seconds for opposite to 11.62 seconds for four access directions.

\subsection{Root node analysis} 
Table  \ref{Tab:results_root_node} provides an analysis of the performance of our approach in the root node. We compare the lower bound heuristics for the root node with the exact result of required moves for all solved instances. For a single tier, the root node gap decreases from 7.89\% for a single access direction to 0.00\% for three and four access directions. In the case of two tiers, the root node gap decreases from 20.95\% to 0.07\% for four access directions. 
This remarkable decrease in gap is likely due to generally fewer required moves, a larger action and solution space, as well as shallow  
virtual lanes. 
In case of two access directions, the lower bound for opposite directions is slightly more accurate than for the corner case. Reason are generally less moves and also more shallow virtual lanes for opposite directions.

\begin{table}[]
\centering
\begin{tabular}{ M{3.00cm}M{3.00cm}M{3.00cm}}
 \hline
 Access Directions & Mean Root Node Gap 1-tier in \% & Mean Root Node Gap 2-tier in \%\\ 
 \hline
 single & 7.89 & 20.95\\
 corner & 0.86 & 3.67\\
 opposite & 0.15 & 3.31\\
 three & 0.00 & 0.94\\
 four & 0.00 & 0.07\\
 \hline
\end{tabular}
\caption{\label{Tab:results_root_node} Root node gaps of our proposed lower bound to the optimal solution.}
\end{table}

\subsection{Required number of moves}
\begin{figure}[]
    \centering
    \includegraphics[width=10cm]{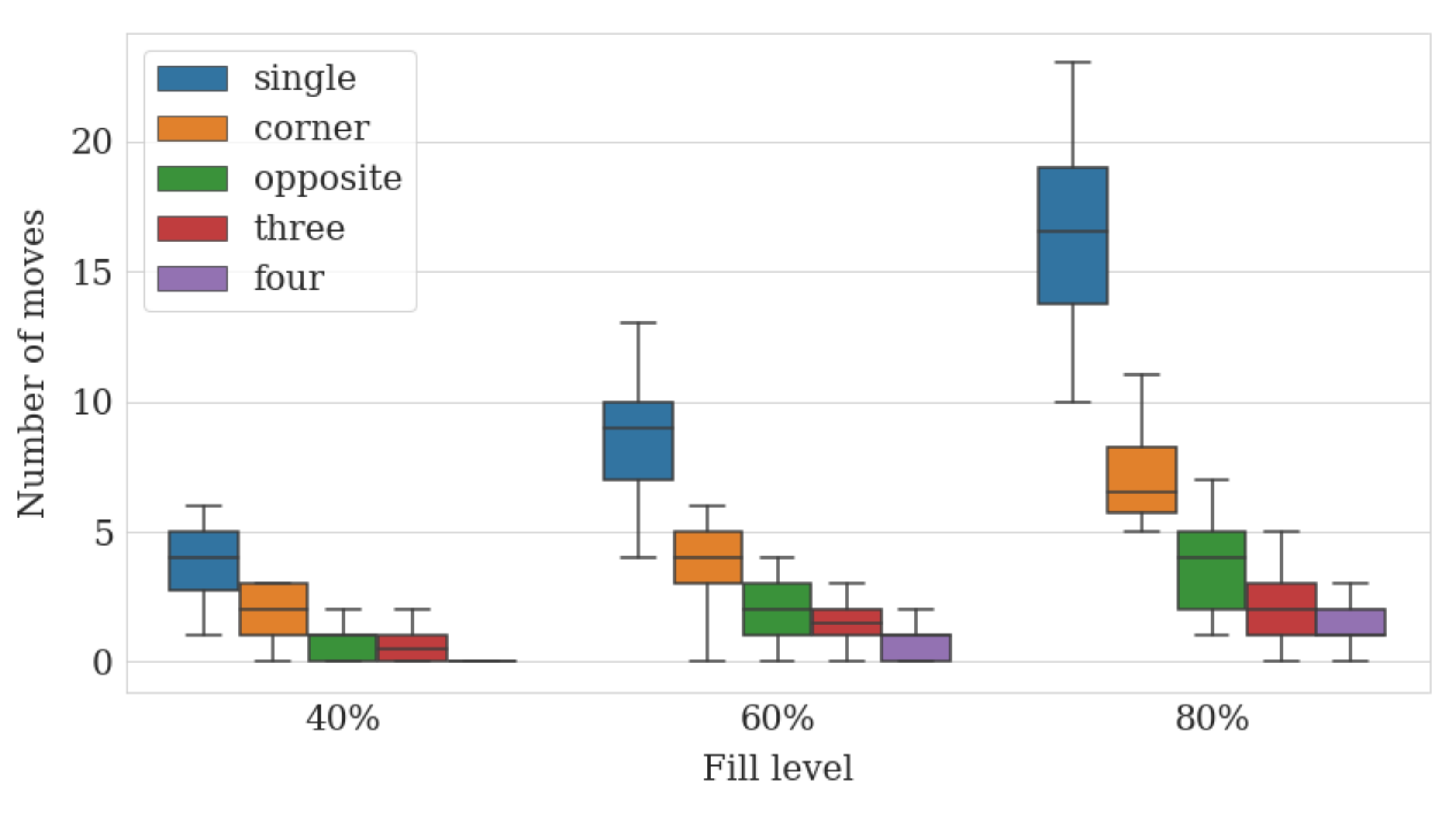}
    \caption{Box plots (without outliers) for size 5x5x1 showing the number of moves for each access variant over all fill levels. This is based on the solved instances shown in Figure \ref{fig:instances_runtime}.}
    \label{fig:moves_boxbplot}
\end{figure}

Figure \ref{fig:moves_boxbplot} shows box plots for the number of required moves for each access direction and all fill levels for size 5x5x1. This is the only size where all instances could be solved without any infeasible instances or timeouts. We provide box plots for all other sizes in Appendix~\ref{appendix_moves}, but note that the general trend of the results are the same as seen here. 
The plots clearly illustrate the big advantage of multiple access directions compared to a single access direction. Unsurprisingly, the number of required moves goes down for each additional access direction that is available, but the amount of decline is significant even for just a single extra access direction. 
Furthermore, high fill levels are especially challenging as they require many moves to sort, but there are not many free storage locations available. 

\begin{figure}[]
    \centering
    \includegraphics[width=10cm]{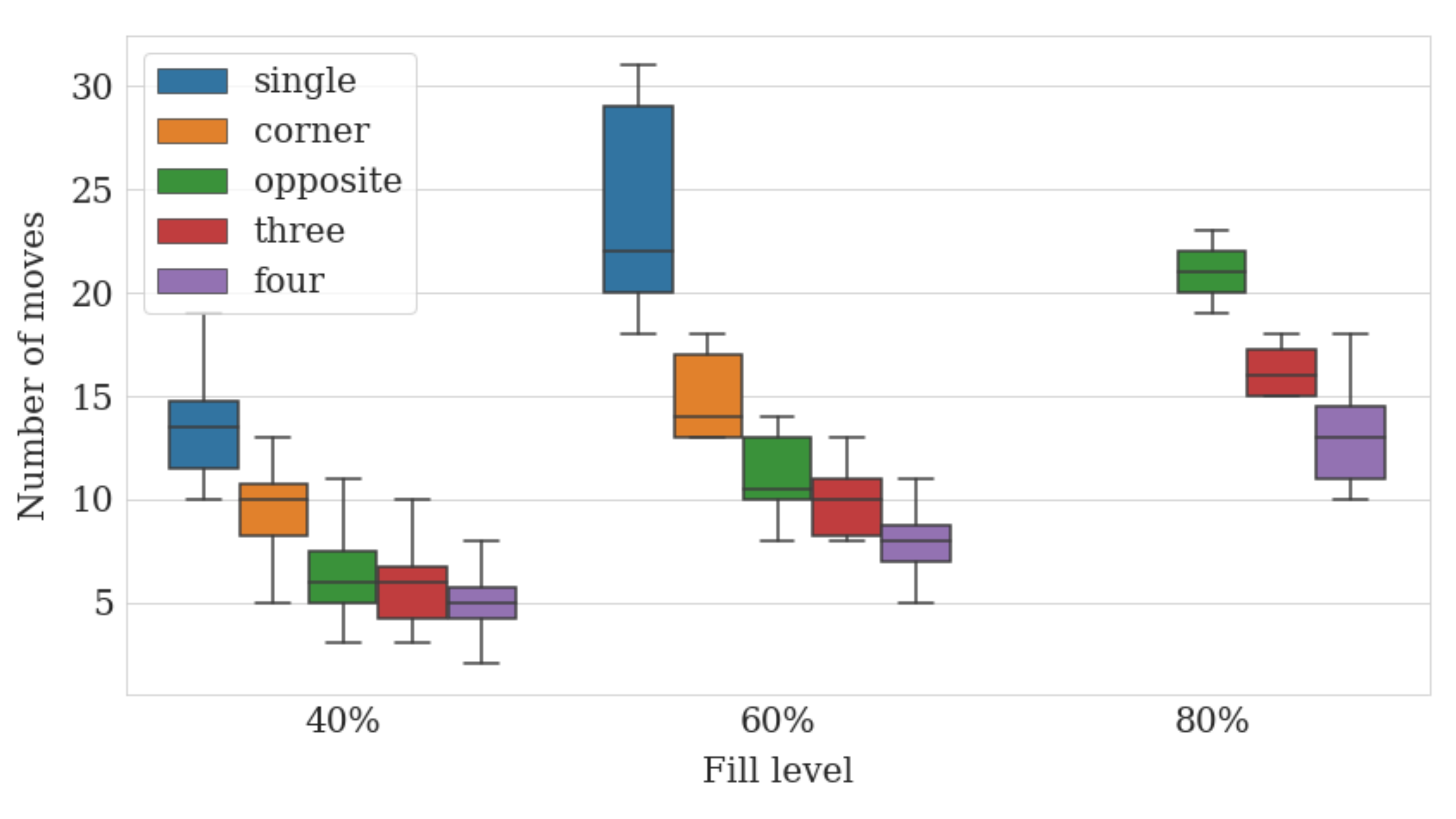}
    \caption{Box plots (without outliers) for size 5x5x2 showing the number of moves for each access variant over all fill levels. This is based on the solved instances shown in Figure \ref{fig:instances_runtime_two_tiers}.
    }
    \label{fig:moves_boxbplot_tier2_5x5}
\end{figure}

The same footprint for the two tier 5x5x2 instances in Figure \ref{fig:moves_boxbplot_tier2_5x5} shows, in general, a similar picture as for a single tier. 
However, in contrast to the single tier instances, not all instances could be solved (see Figure \ref{fig:instances_runtime_two_tiers}). 
The main issue with multiple tiers is that the likelihood of encountering an unsorted stack grows with the size of the instance, and this can require several unit load movements to fix before the rest of the stacks can be dealt with.

\section{Impact of multiple access directions for operations}
\label{7_insights}

The main advantage of block stacking storage systems under a fully shared storage policy is to achieve high storage densities. 
However, high storage density comes at the cost of decreased accessibility due to blockage.
Our results clearly show that this disadvantage can be significantly mitigated if up to four access directions are supported both technically by the load carriers and algorithmically in the control system.
As discussed in the following, the number of access directions immensely impact the system in three  dimensions of (1) efficiency and feasibility, (2) layout decisions, and (3) traffic management.

\paragraph{Efficiency and feasibility}
Our results show that the required number of moves to sort a storage bay drop considerably with an increasing number of access directions. This becomes especially visible at high fill levels. 
The main reasons are (1) an increased number of directly accessible unit loads, (2) a decrease in average lane depth (less blocking unit loads and better free slots), and (3) variable determination of 
lane depth and configuration (increased flexibility to access unit loads).


Our experiments show clearly that only using a single access leads to a large number of infeasible instances at high fill rates. 
Thus, single access is more suitable for scenarios where no sorting is necessary.  
Increasing the number of tiers also leads to a much higher number of moves, as well as many infeasible instances. Hence, we recommend to allow mixing of SKUs in lanes but not within a stack (pure stacks). 


\paragraph{Layout decisions}
Access directions have a direct impact on 
layout decisions. Indeed, our study results clearly suggest that when planning an aisle configuration, planners should consider the use of multiple access directions. 
Of course, this results in a trade-off between aisle- and storage-space (fast access and travel times versus a high space utilization). Further access directions might require additional aisle space, but can enable deeper bays without suffering bad access times. 
Another aspect to be considered during layout decisions are the different results for two sided access: If possible, an access from two opposite directions is preferable to corner access as the pre-marshalling results in less moves. Reason is that in the opposite setup lanes are always divided in two parts and, hence, are shorter on average. Furthermore, in an opposite access direction setup for vehicle operations no access from crossing directions needs to be considered and most standard load carriers support opposite access (e.g., pallets).

\paragraph{Operational control}

Arrival and retrieval times of unit loads can be stochastic. For scenarios with high uncertainty for the in- and outbound flow sequence, the higher number of directly accessible unit loads and the higher number of access direction options in a multiple access scenario is advantageous as less sorting effort results. 

Furthermore, the concept of virtual lanes is beneficial for the vehicle management.
As we build only contiguous virtual lanes, no crossing paths of vehicles---as illustrated in the examples of Section~\ref{sec:impact_access_direction_fixing}---can occur.
Hence, vehicle management does not need to consider collision avoidance  within the bay.\\

In general, our results indicate that multiple access directions can greatly increase sorting efficiency for high-density systems, and we suspect that this will hold true also when for storage and retrieval operations. However, further investigations are needed to confirm the value of multiple access directions in a more general intralogistics setting.

\section{Conclusion and future research}
\label{8_conlusion}
We introduce the UPMP for sorting block stacking storage systems in idle time. 
Solving the UPMP allows organizations to increase the storage density of their block stacking storage, while still ensuring fast access times. This enables new kinds of compact layouts for many types of storage systems that consist of movable floor units, like pallet-based block storage, RMFS, container yards, material transport trolleys and many more. 

Our solution approach sorts storage bays with any combination of up to four access directions to minimize the number of moves. 
In the case of the UPMP with a single access direction, we show that the solution approach from the CPMP can be directly applied.
For any combination of multiple access directions, we present a novel two-step approach that fixes the access directions for each stack throughout the 
pre-marshalling process and allows for solving real-world sized instances with up to 81 storage locations without any timeouts within a limit of one hour. 

The experiments show that multiple access directions significantly reduce the required number of moves and allow solving much more instances with faster runtimes, especially for high fill levels. Our detailed comparison of access directions and tiers provides valuable insights for the development and layout design of highly dense block storage systems.

In a next step, several future research directions are possible. Our search procedure and adapted lower bound heuristic proves to be strong with a very low root node gap for multiple access directions. 
However, further improvements to accelerate the search are possible 
(see e.g., \citet{tierneySolvingPremarshallingProblem2017, tanakaSolvingRealworldSized2018, tanakaBranchBoundApproach2019}). 





Furthermore, future work is required to increase the applicability of our approach to a variety of use cases. 
Extending the scope to large, multi-bay warehouses would require adjusting the objective of minimizing the number of moves to also consider travel time or distance (e.g., \citet{parreno2020minimizing}). Finally, heuristic approaches have had significant success for the CPMP, and could be also useful for solving large instances of the UPMP. 

\bibliographystyle{apacite}
\bibliography{Literature.bib}

\begin{thebibliography}{}

\bibitem [\protect \citeauthoryear {%
Ball%
}{%
Ball%
}{%
{\protect \APACyear {1893}}%
}]{%
ballMathematicalRecreationsEssays1893}
\APACinsertmetastar {%
ballMathematicalRecreationsEssays1893}%
\begin{APACrefauthors}%
Ball, W\BPBI R.%
\end{APACrefauthors}%
\unskip\
\newblock
\APACrefYearMonthDay{1893}{}{}.
\newblock
{\BBOQ}\APACrefatitle {Mathematical {Recreations} and {Essays}} {Mathematical
  {Recreations} and {Essays}}.{\BBCQ}
\newblock
\APACjournalVolNumPages{Bulletin des sciences mathématiques}{17}{}{105--107}.
\PrintBackRefs{\CurrentBib}

\bibitem [\protect \citeauthoryear {%
Bortfeldt%
\ \BBA {} Forster%
}{%
Bortfeldt%
\ \BBA {} Forster%
}{%
{\protect \APACyear {2012}}%
}]{%
bortfeldtTreeSearchProcedure2012}
\APACinsertmetastar {%
bortfeldtTreeSearchProcedure2012}%
\begin{APACrefauthors}%
Bortfeldt, A.%
\BCBT {}\ \BBA {} Forster, F.%
\end{APACrefauthors}%
\unskip\
\newblock
\APACrefYearMonthDay{2012}{}{}.
\newblock
{\BBOQ}\APACrefatitle {A tree search procedure for the container
  pre-marshalling problem} {A tree search procedure for the container
  pre-marshalling problem}.{\BBCQ}
\newblock
\APACjournalVolNumPages{European Journal of Operational
  Research}{217}{3}{531--540}.
\PrintBackRefs{\CurrentBib}

\bibitem [\protect \citeauthoryear {%
Boysen%
, Boywitz%
\BCBL {}\ \BBA {} Weidinger%
}{%
Boysen%
\ \protect \BOthers {.}}{%
{\protect \APACyear {2018}}%
}]{%
boysen2018deep}
\APACinsertmetastar {%
boysen2018deep}%
\begin{APACrefauthors}%
Boysen, N.%
, Boywitz, D.%
\BCBL {}\ \BBA {} Weidinger, F.%
\end{APACrefauthors}%
\unskip\
\newblock
\APACrefYearMonthDay{2018}{}{}.
\newblock
{\BBOQ}\APACrefatitle {Deep-lane storage of time-critical items: one-sided
  versus two-sided access} {Deep-lane storage of time-critical items: one-sided
  versus two-sided access}.{\BBCQ}
\newblock
\APACjournalVolNumPages{OR Spectrum}{40}{4}{1141--1170}.
\PrintBackRefs{\CurrentBib}

\bibitem [\protect \citeauthoryear {%
Caserta%
, Schwarze%
\BCBL {}\ \BBA {} Voß%
}{%
Caserta%
\ \protect \BOthers {.}}{%
{\protect \APACyear {2011}}%
}]{%
casertaContainerRehandlingMaritime2011}
\APACinsertmetastar {%
casertaContainerRehandlingMaritime2011}%
\begin{APACrefauthors}%
Caserta, M.%
, Schwarze, S.%
\BCBL {}\ \BBA {} Voß, S.%
\end{APACrefauthors}%
\unskip\
\newblock
\APACrefYearMonthDay{2011}{}{}.
\newblock
{\BBOQ}\APACrefatitle {Container {Rehandling} at {Maritime} {Container}
  {Terminals}} {Container {Rehandling} at {Maritime} {Container}
  {Terminals}}.{\BBCQ}
\newblock
\BIn{} J\BPBI W.~Böse\ (\BED), \APACrefbtitle {Handbook of {Terminal}
  {Planning}} {Handbook of {Terminal} {Planning}}\ (\BPGS\ 247--269).
\PrintBackRefs{\CurrentBib}

\bibitem [\protect \citeauthoryear {%
Covic%
}{%
Covic%
}{%
{\protect \APACyear {2017}}%
}]{%
covicRemarshallingAutomatedContainer2017}
\APACinsertmetastar {%
covicRemarshallingAutomatedContainer2017}%
\begin{APACrefauthors}%
Covic, F.%
\end{APACrefauthors}%
\unskip\
\newblock
\APACrefYearMonthDay{2017}{}{}.
\newblock
{\BBOQ}\APACrefatitle {Re-marshalling in automated container yards with
  terminal appointment systems} {Re-marshalling in automated container yards
  with terminal appointment systems}.{\BBCQ}
\newblock
\APACjournalVolNumPages{Flexible Services and Manufacturing
  Journal}{29}{3}{433--503}.
\PrintBackRefs{\CurrentBib}

\bibitem [\protect \citeauthoryear {%
Expósito-Izquierdo%
, Melián-Batista%
\BCBL {}\ \BBA {} Moreno-Vega%
}{%
Expósito-Izquierdo%
\ \protect \BOthers {.}}{%
{\protect \APACyear {2012}}%
}]{%
exposito-izquierdoPreMarshallingProblemHeuristic2012}
\APACinsertmetastar {%
exposito-izquierdoPreMarshallingProblemHeuristic2012}%
\begin{APACrefauthors}%
Expósito-Izquierdo, C.%
, Melián-Batista, B.%
\BCBL {}\ \BBA {} Moreno-Vega, M.%
\end{APACrefauthors}%
\unskip\
\newblock
\APACrefYearMonthDay{2012}{}{}.
\newblock
{\BBOQ}\APACrefatitle {Pre-{Marshalling} {Problem}: {Heuristic} solution method
  and instances generator} {Pre-{Marshalling} {Problem}: {Heuristic} solution
  method and instances generator}.{\BBCQ}
\newblock
\APACjournalVolNumPages{Expert Systems with Applications}{39}{9}{8337--8349}.
\PrintBackRefs{\CurrentBib}

\bibitem [\protect \citeauthoryear {%
Fragapane%
, De~Koster%
, Sgarbossa%
\BCBL {}\ \BBA {} Strandhagen%
}{%
Fragapane%
\ \protect \BOthers {.}}{%
{\protect \APACyear {2021}}%
}]{%
fragapane2021planning}
\APACinsertmetastar {%
fragapane2021planning}%
\begin{APACrefauthors}%
Fragapane, G.%
, De~Koster, R.%
, Sgarbossa, F.%
\BCBL {}\ \BBA {} Strandhagen, J\BPBI O.%
\end{APACrefauthors}%
\unskip\
\newblock
\APACrefYearMonthDay{2021}{}{}.
\newblock
{\BBOQ}\APACrefatitle {Planning and control of autonomous mobile robots for
  intralogistics: Literature review and research agenda} {Planning and control
  of autonomous mobile robots for intralogistics: Literature review and
  research agenda}.{\BBCQ}
\newblock
\APACjournalVolNumPages{European Journal of Operational
  Research}{294}{2}{405--426}.
\PrintBackRefs{\CurrentBib}

\bibitem [\protect \citeauthoryear {%
Ge%
, Meng%
, Liu%
, Tang%
\BCBL {}\ \BBA {} Zhao%
}{%
Ge%
\ \protect \BOthers {.}}{%
{\protect \APACyear {2020}}%
}]{%
geLogisticsOptimisationSlab2020}
\APACinsertmetastar {%
geLogisticsOptimisationSlab2020}%
\begin{APACrefauthors}%
Ge, P.%
, Meng, Y.%
, Liu, J.%
, Tang, L.%
\BCBL {}\ \BBA {} Zhao, R.%
\end{APACrefauthors}%
\unskip\
\newblock
\APACrefYearMonthDay{2020}{}{}.
\newblock
{\BBOQ}\APACrefatitle {Logistics optimisation of slab pre-marshalling problem
  in steel industry} {Logistics optimisation of slab pre-marshalling problem in
  steel industry}.{\BBCQ}
\newblock
\APACjournalVolNumPages{International Journal of Production
  Research}{58}{13}{4050--4070}.
\PrintBackRefs{\CurrentBib}

\bibitem [\protect \citeauthoryear {%
Ge%
, Zhao%
, Sun%
\BCBL {}\ \BBA {} Dong%
}{%
Ge%
\ \protect \BOthers {.}}{%
{\protect \APACyear {2021}}%
}]{%
geIntegratedOptimisationStorage2021a}
\APACinsertmetastar {%
geIntegratedOptimisationStorage2021a}%
\begin{APACrefauthors}%
Ge, P.%
, Zhao, R.%
, Sun, D.%
\BCBL {}\ \BBA {} Dong, Y.%
\end{APACrefauthors}%
\unskip\
\newblock
\APACrefYearMonthDay{2021}{}{}.
\newblock
{\BBOQ}\APACrefatitle {Integrated optimisation of storage and pre-marshalling
  moves in a slab warehouse} {Integrated optimisation of storage and
  pre-marshalling moves in a slab warehouse}.{\BBCQ}
\newblock
\APACjournalVolNumPages{International Journal of Production
  Research}{0}{0}{1--23}.
\PrintBackRefs{\CurrentBib}

\bibitem [\protect \citeauthoryear {%
Goetschalckx%
\ \BBA {} Ratliff%
}{%
Goetschalckx%
\ \BBA {} Ratliff%
}{%
{\protect \APACyear {1990}}%
}]{%
goetschalckx1990shared}
\APACinsertmetastar {%
goetschalckx1990shared}%
\begin{APACrefauthors}%
Goetschalckx, M.%
\BCBT {}\ \BBA {} Ratliff, H\BPBI D.%
\end{APACrefauthors}%
\unskip\
\newblock
\APACrefYearMonthDay{1990}{}{}.
\newblock
{\BBOQ}\APACrefatitle {Shared storage policies based on the duration stay of
  unit loads} {Shared storage policies based on the duration stay of unit
  loads}.{\BBCQ}
\newblock
\APACjournalVolNumPages{Management science}{36}{9}{1120--1132}.
\PrintBackRefs{\CurrentBib}

\bibitem [\protect \citeauthoryear {%
Gue%
\ \BBA {} Kim%
}{%
Gue%
\ \BBA {} Kim%
}{%
{\protect \APACyear {2007}}%
}]{%
guePuzzlebasedStorageSystems2007}
\APACinsertmetastar {%
guePuzzlebasedStorageSystems2007}%
\begin{APACrefauthors}%
Gue, K\BPBI R.%
\BCBT {}\ \BBA {} Kim, B\BPBI S.%
\end{APACrefauthors}%
\unskip\
\newblock
\APACrefYearMonthDay{2007}{}{}.
\newblock
{\BBOQ}\APACrefatitle {Puzzle-based storage systems} {Puzzle-based storage
  systems}.{\BBCQ}
\newblock
\APACjournalVolNumPages{Naval Research Logistics (NRL)}{54}{5}{556--567}.
\PrintBackRefs{\CurrentBib}

\bibitem [\protect \citeauthoryear {%
Hart%
, Nilsson%
\BCBL {}\ \BBA {} Raphael%
}{%
Hart%
\ \protect \BOthers {.}}{%
{\protect \APACyear {1968}}%
}]{%
hart1968formal}
\APACinsertmetastar {%
hart1968formal}%
\begin{APACrefauthors}%
Hart, P\BPBI E.%
, Nilsson, N\BPBI J.%
\BCBL {}\ \BBA {} Raphael, B.%
\end{APACrefauthors}%
\unskip\
\newblock
\APACrefYearMonthDay{1968}{}{}.
\newblock
{\BBOQ}\APACrefatitle {A formal basis for the heuristic determination of
  minimum cost paths} {A formal basis for the heuristic determination of
  minimum cost paths}.{\BBCQ}
\newblock
\APACjournalVolNumPages{IEEE transactions on Systems Science and
  Cybernetics}{4}{2}{100--107}.
\PrintBackRefs{\CurrentBib}

\bibitem [\protect \citeauthoryear {%
Jin%
, Yang%
\BCBL {}\ \BBA {} Duan%
}{%
Jin%
\ \protect \BOthers {.}}{%
{\protect \APACyear {2020}}%
}]{%
jinMultipleDeepLayout2020a}
\APACinsertmetastar {%
jinMultipleDeepLayout2020a}%
\begin{APACrefauthors}%
Jin, G.%
, Yang, P.%
\BCBL {}\ \BBA {} Duan, G.%
\end{APACrefauthors}%
\unskip\
\newblock
\APACrefYearMonthDay{2020}{}{}.
\newblock
{\BBOQ}\APACrefatitle {Multiple {Deep} {Layout} of {Robotic} {Mobile}
  {Fulfillment} {System}} {Multiple {Deep} {Layout} of {Robotic} {Mobile}
  {Fulfillment} {System}}.{\BBCQ}
\newblock
\BIn{} \APACrefbtitle {2020 {IEEE} 7th {International} {Conference} on
  {Industrial} {Engineering} and {Applications} ({ICIEA})} {2020 {IEEE} 7th
  {International} {Conference} on {Industrial} {Engineering} and {Applications}
  ({ICIEA})}\ (\BPGS\ 230--234).
\PrintBackRefs{\CurrentBib}

\bibitem [\protect \citeauthoryear {%
König%
, Lübbecke%
, Möhring%
, Schäfer%
\BCBL {}\ \BBA {} Spenke%
}{%
König%
\ \protect \BOthers {.}}{%
{\protect \APACyear {2007}}%
}]{%
konigSolutionsRealWorldInstances2007}
\APACinsertmetastar {%
konigSolutionsRealWorldInstances2007}%
\begin{APACrefauthors}%
König, F\BPBI G.%
, Lübbecke, M.%
, Möhring, R.%
, Schäfer, G.%
\BCBL {}\ \BBA {} Spenke, I.%
\end{APACrefauthors}%
\unskip\
\newblock
\APACrefYearMonthDay{2007}{}{}.
\newblock
{\BBOQ}\APACrefatitle {Solutions to {Real}-{World} {Instances} of
  {PSPACE}-{Complete} {Stacking}} {Solutions to {Real}-{World} {Instances} of
  {PSPACE}-{Complete} {Stacking}}.{\BBCQ}
\newblock
\BIn{} L.~Arge, M.~Hoffmann\BCBL {}\ \BBA {} E.~Welzl\ (\BEDS), \APACrefbtitle
  {Algorithms – {ESA} 2007} {Algorithms – {ESA} 2007}\ (\BPGS\ 729--740).
\PrintBackRefs{\CurrentBib}

\bibitem [\protect \citeauthoryear {%
Lee%
\ \BBA {} Chao%
}{%
Lee%
\ \BBA {} Chao%
}{%
{\protect \APACyear {2009}}%
}]{%
LEE2009468}
\APACinsertmetastar {%
LEE2009468}%
\begin{APACrefauthors}%
Lee, Y.%
\BCBT {}\ \BBA {} Chao, S\BHBI L.%
\end{APACrefauthors}%
\unskip\
\newblock
\APACrefYearMonthDay{2009}{}{}.
\newblock
{\BBOQ}\APACrefatitle {A neighborhood search heuristic for pre-marshalling
  export containers} {A neighborhood search heuristic for pre-marshalling
  export containers}.{\BBCQ}
\newblock
\APACjournalVolNumPages{European Journal of Operational
  Research}{196}{2}{468-475}.
\PrintBackRefs{\CurrentBib}

\bibitem [\protect \citeauthoryear {%
Lee%
\ \BBA {} Hsu%
}{%
Lee%
\ \BBA {} Hsu%
}{%
{\protect \APACyear {2007}}%
}]{%
lee2007optimization}
\APACinsertmetastar {%
lee2007optimization}%
\begin{APACrefauthors}%
Lee, Y.%
\BCBT {}\ \BBA {} Hsu, N\BHBI Y.%
\end{APACrefauthors}%
\unskip\
\newblock
\APACrefYearMonthDay{2007}{}{}.
\newblock
{\BBOQ}\APACrefatitle {An optimization model for the container pre-marshalling
  problem} {An optimization model for the container pre-marshalling
  problem}.{\BBCQ}
\newblock
\APACjournalVolNumPages{Computers \& operations research}{34}{11}{3295--3313}.
\PrintBackRefs{\CurrentBib}

\bibitem [\protect \citeauthoryear {%
Lehnfeld%
\ \BBA {} Knust%
}{%
Lehnfeld%
\ \BBA {} Knust%
}{%
{\protect \APACyear {2014}}%
}]{%
lehnfeldLoadingUnloadingPremarshalling2014}
\APACinsertmetastar {%
lehnfeldLoadingUnloadingPremarshalling2014}%
\begin{APACrefauthors}%
Lehnfeld, J.%
\BCBT {}\ \BBA {} Knust, S.%
\end{APACrefauthors}%
\unskip\
\newblock
\APACrefYearMonthDay{2014}{}{}.
\newblock
{\BBOQ}\APACrefatitle {Loading, unloading and premarshalling of stacks in
  storage areas: {Survey} and classification} {Loading, unloading and
  premarshalling of stacks in storage areas: {Survey} and
  classification}.{\BBCQ}
\newblock
\APACjournalVolNumPages{European Journal of Operational
  Research}{239}{2}{297--312}.
\PrintBackRefs{\CurrentBib}

\bibitem [\protect \citeauthoryear {%
Maniezzo%
, Boschetti%
\BCBL {}\ \BBA {} Gutjahr%
}{%
Maniezzo%
\ \protect \BOthers {.}}{%
{\protect \APACyear {2021}}%
}]{%
maniezzoStochasticPremarshallingBlock2021}
\APACinsertmetastar {%
maniezzoStochasticPremarshallingBlock2021}%
\begin{APACrefauthors}%
Maniezzo, V.%
, Boschetti, M\BPBI A.%
\BCBL {}\ \BBA {} Gutjahr, W\BPBI J.%
\end{APACrefauthors}%
\unskip\
\newblock
\APACrefYearMonthDay{2021}{}{}.
\newblock
{\BBOQ}\APACrefatitle {Stochastic premarshalling of block stacking warehouses}
  {Stochastic premarshalling of block stacking warehouses}.{\BBCQ}
\newblock
\APACjournalVolNumPages{Omega}{102}{}{102336}.
\PrintBackRefs{\CurrentBib}

\bibitem [\protect \citeauthoryear {%
Merschformann%
}{%
Merschformann%
}{%
{\protect \APACyear {2018}}%
}]{%
merschformannActiveRepositioningStorage2018}
\APACinsertmetastar {%
merschformannActiveRepositioningStorage2018}%
\begin{APACrefauthors}%
Merschformann, M.%
\end{APACrefauthors}%
\unskip\
\newblock
\APACrefYearMonthDay{2018}{}{}.
\newblock
{\BBOQ}\APACrefatitle {Active {Repositioning} of {Storage} {Units} in {Robotic}
  {Mobile} {Fulfillment} {Systems}} {Active {Repositioning} of {Storage}
  {Units} in {Robotic} {Mobile} {Fulfillment} {Systems}}.{\BBCQ}
\newblock
\BIn{} N.~Kliewer, J\BPBI F.~Ehmke\BCBL {}\ \BBA {} R.~Borndörfer\ (\BEDS),
  \APACrefbtitle {Operations {Research} {Proceedings} 2017} {Operations
  {Research} {Proceedings} 2017}\ (\BPGS\ 379--385).
\PrintBackRefs{\CurrentBib}

\bibitem [\protect \citeauthoryear {%
Parre{\~n}o-Torres%
, Alvarez-Valdes%
, Ruiz%
\BCBL {}\ \BBA {} Tierney%
}{%
Parre{\~n}o-Torres%
\ \protect \BOthers {.}}{%
{\protect \APACyear {2020}}%
}]{%
parreno2020minimizing}
\APACinsertmetastar {%
parreno2020minimizing}%
\begin{APACrefauthors}%
Parre{\~n}o-Torres, C.%
, Alvarez-Valdes, R.%
, Ruiz, R.%
\BCBL {}\ \BBA {} Tierney, K.%
\end{APACrefauthors}%
\unskip\
\newblock
\APACrefYearMonthDay{2020}{}{}.
\newblock
{\BBOQ}\APACrefatitle {Minimizing crane times in pre-marshalling problems}
  {Minimizing crane times in pre-marshalling problems}.{\BBCQ}
\newblock
\APACjournalVolNumPages{Transportation Research Part E: Logistics and
  Transportation Review}{137}{}{101917}.
\PrintBackRefs{\CurrentBib}

\bibitem [\protect \citeauthoryear {%
Parreño-Torres%
, Alvarez-Valdes%
\BCBL {}\ \BBA {} Ruiz%
}{%
Parreño-Torres%
\ \protect \BOthers {.}}{%
{\protect \APACyear {2019}}%
}]{%
parreno-torresIntegerProgrammingModels2019}
\APACinsertmetastar {%
parreno-torresIntegerProgrammingModels2019}%
\begin{APACrefauthors}%
Parreño-Torres, C.%
, Alvarez-Valdes, R.%
\BCBL {}\ \BBA {} Ruiz, R.%
\end{APACrefauthors}%
\unskip\
\newblock
\APACrefYearMonthDay{2019}{}{}.
\newblock
{\BBOQ}\APACrefatitle {Integer programming models for the pre-marshalling
  problem} {Integer programming models for the pre-marshalling problem}.{\BBCQ}
\newblock
\APACjournalVolNumPages{European Journal of Operational
  Research}{274}{1}{142--154}.
\PrintBackRefs{\CurrentBib}

\bibitem [\protect \citeauthoryear {%
Pfrommer%
\ \BBA {} Meyer%
}{%
Pfrommer%
\ \BBA {} Meyer%
}{%
{\protect \APACyear {2020}}%
}]{%
pfrommer2020autonomously}
\APACinsertmetastar {%
pfrommer2020autonomously}%
\begin{APACrefauthors}%
Pfrommer, J.%
\BCBT {}\ \BBA {} Meyer, A.%
\end{APACrefauthors}%
\unskip\
\newblock
\APACrefYearMonthDay{2020}{}{}.
\newblock
{\BBOQ}\APACrefatitle {Autonomously organized block stacking warehouses: A
  review of decision problems and major challenges} {Autonomously organized
  block stacking warehouses: A review of decision problems and major
  challenges}.{\BBCQ}
\newblock
\APACjournalVolNumPages{Logistics Journal: Proceedings}{2020}{12}{}.
\PrintBackRefs{\CurrentBib}

\bibitem [\protect \citeauthoryear {%
Prandtstetter%
}{%
Prandtstetter%
}{%
{\protect \APACyear {2013}}%
}]{%
prandtstetter2013dynamic}
\APACinsertmetastar {%
prandtstetter2013dynamic}%
\begin{APACrefauthors}%
Prandtstetter, M.%
\end{APACrefauthors}%
\unskip\
\newblock
\APACrefYearMonthDay{2013}{}{}.
\newblock
{\BBOQ}\APACrefatitle {A dynamic programming based branch-and-bound algorithm
  for the container pre-marshalling problem} {A dynamic programming based
  branch-and-bound algorithm for the container pre-marshalling problem}.{\BBCQ}
\newblock
\APACjournalVolNumPages{Technical repot, IT Austrian institute of
  technology}{}{}{}.
\PrintBackRefs{\CurrentBib}

\bibitem [\protect \citeauthoryear {%
Rendl%
\ \BBA {} Prandtstetter%
}{%
Rendl%
\ \BBA {} Prandtstetter%
}{%
{\protect \APACyear {2013}}%
}]{%
rendlConstraintModelsContainer2013}
\APACinsertmetastar {%
rendlConstraintModelsContainer2013}%
\begin{APACrefauthors}%
Rendl, A.%
\BCBT {}\ \BBA {} Prandtstetter, M.%
\end{APACrefauthors}%
\unskip\
\newblock
\APACrefYearMonthDay{2013}{}{}.
\newblock
{\BBOQ}\APACrefatitle {Constraint models for the container pre-marshaling
  problem} {Constraint models for the container pre-marshaling problem}.{\BBCQ}
\newblock
\APACjournalVolNumPages{ModRef}{2013}{}{12th}.
\PrintBackRefs{\CurrentBib}

\bibitem [\protect \citeauthoryear {%
Tanaka%
\ \BBA {} Tierney%
}{%
Tanaka%
\ \BBA {} Tierney%
}{%
{\protect \APACyear {2018}}%
}]{%
tanakaSolvingRealworldSized2018}
\APACinsertmetastar {%
tanakaSolvingRealworldSized2018}%
\begin{APACrefauthors}%
Tanaka, S.%
\BCBT {}\ \BBA {} Tierney, K.%
\end{APACrefauthors}%
\unskip\
\newblock
\APACrefYearMonthDay{2018}{}{}.
\newblock
{\BBOQ}\APACrefatitle {Solving real-world sized container pre-marshalling
  problems with an iterative deepening branch-and-bound algorithm} {Solving
  real-world sized container pre-marshalling problems with an iterative
  deepening branch-and-bound algorithm}.{\BBCQ}
\newblock
\APACjournalVolNumPages{European Journal of Operational
  Research}{264}{1}{165--180}.
\PrintBackRefs{\CurrentBib}

\bibitem [\protect \citeauthoryear {%
Tanaka%
, Tierney%
, Parreño-Torres%
, Alvarez-Valdes%
\BCBL {}\ \BBA {} Ruiz%
}{%
Tanaka%
\ \protect \BOthers {.}}{%
{\protect \APACyear {2019}}%
}]{%
tanakaBranchBoundApproach2019}
\APACinsertmetastar {%
tanakaBranchBoundApproach2019}%
\begin{APACrefauthors}%
Tanaka, S.%
, Tierney, K.%
, Parreño-Torres, C.%
, Alvarez-Valdes, R.%
\BCBL {}\ \BBA {} Ruiz, R.%
\end{APACrefauthors}%
\unskip\
\newblock
\APACrefYearMonthDay{2019}{{\APACmonth{10}}}{}.
\newblock
{\BBOQ}\APACrefatitle {A branch and bound approach for large pre-marshalling
  problems} {A branch and bound approach for large pre-marshalling
  problems}.{\BBCQ}
\newblock
\APACjournalVolNumPages{European Journal of Operational
  Research}{278}{1}{211--225}.
\PrintBackRefs{\CurrentBib}

\bibitem [\protect \citeauthoryear {%
Tang%
, Liu%
, Rong%
\BCBL {}\ \BBA {} Yang%
}{%
Tang%
\ \protect \BOthers {.}}{%
{\protect \APACyear {2002}}%
}]{%
tangModellingGeneticAlgorithm2002}
\APACinsertmetastar {%
tangModellingGeneticAlgorithm2002}%
\begin{APACrefauthors}%
Tang, L.%
, Liu, J.%
, Rong, A.%
\BCBL {}\ \BBA {} Yang, Z.%
\end{APACrefauthors}%
\unskip\
\newblock
\APACrefYearMonthDay{2002}{}{}.
\newblock
{\BBOQ}\APACrefatitle {Modelling and a genetic algorithm solution for the slab
  stack shuffling problem when implementing steel rolling schedules} {Modelling
  and a genetic algorithm solution for the slab stack shuffling problem when
  implementing steel rolling schedules}.{\BBCQ}
\newblock
\APACjournalVolNumPages{International Journal of Production
  Research}{40}{7}{1583--1595}.
\PrintBackRefs{\CurrentBib}

\bibitem [\protect \citeauthoryear {%
Tang%
, Liu%
, Yang%
, Li%
\BCBL {}\ \BBA {} Li%
}{%
Tang%
\ \protect \BOthers {.}}{%
{\protect \APACyear {2015}}%
}]{%
tangModelingSolutionShip2015}
\APACinsertmetastar {%
tangModelingSolutionShip2015}%
\begin{APACrefauthors}%
Tang, L.%
, Liu, J.%
, Yang, F.%
, Li, F.%
\BCBL {}\ \BBA {} Li, K.%
\end{APACrefauthors}%
\unskip\
\newblock
\APACrefYearMonthDay{2015}{}{}.
\newblock
{\BBOQ}\APACrefatitle {Modeling and solution for the ship stowage planning
  problem of coils in the steel industry} {Modeling and solution for the ship
  stowage planning problem of coils in the steel industry}.{\BBCQ}
\newblock
\APACjournalVolNumPages{Naval Research Logistics (NRL)}{62}{7}{564--581}.
\PrintBackRefs{\CurrentBib}

\bibitem [\protect \citeauthoryear {%
Tang%
, Zhao%
\BCBL {}\ \BBA {} Liu%
}{%
Tang%
\ \protect \BOthers {.}}{%
{\protect \APACyear {2012}}%
}]{%
tangModelsAlgorithmsShuffling2012}
\APACinsertmetastar {%
tangModelsAlgorithmsShuffling2012}%
\begin{APACrefauthors}%
Tang, L.%
, Zhao, R.%
\BCBL {}\ \BBA {} Liu, J.%
\end{APACrefauthors}%
\unskip\
\newblock
\APACrefYearMonthDay{2012}{}{}.
\newblock
{\BBOQ}\APACrefatitle {Models and algorithms for shuffling problems in steel
  plants} {Models and algorithms for shuffling problems in steel
  plants}.{\BBCQ}
\newblock
\APACjournalVolNumPages{Naval Research Logistics (NRL)}{59}{7}{502--524}.
\PrintBackRefs{\CurrentBib}

\bibitem [\protect \citeauthoryear {%
Tierney%
, Pacino%
\BCBL {}\ \BBA {} Voß%
}{%
Tierney%
\ \protect \BOthers {.}}{%
{\protect \APACyear {2017}}%
}]{%
tierneySolvingPremarshallingProblem2017}
\APACinsertmetastar {%
tierneySolvingPremarshallingProblem2017}%
\begin{APACrefauthors}%
Tierney, K.%
, Pacino, D.%
\BCBL {}\ \BBA {} Voß, S.%
\end{APACrefauthors}%
\unskip\
\newblock
\APACrefYearMonthDay{2017}{{\APACmonth{06}}}{}.
\newblock
{\BBOQ}\APACrefatitle {Solving the pre-marshalling problem to optimality with
  {A}* and {IDA}*} {Solving the pre-marshalling problem to optimality with {A}*
  and {IDA}*}.{\BBCQ}
\newblock
\APACjournalVolNumPages{Flexible Services and Manufacturing
  Journal}{29}{2}{223--259}.
\PrintBackRefs{\CurrentBib}

\bibitem [\protect \citeauthoryear {%
Wurman%
, D'Andrea%
\BCBL {}\ \BBA {} Mountz%
}{%
Wurman%
\ \protect \BOthers {.}}{%
{\protect \APACyear {2008}}%
}]{%
wurman2008coordinating}
\APACinsertmetastar {%
wurman2008coordinating}%
\begin{APACrefauthors}%
Wurman, P\BPBI R.%
, D'Andrea, R.%
\BCBL {}\ \BBA {} Mountz, M.%
\end{APACrefauthors}%
\unskip\
\newblock
\APACrefYearMonthDay{2008}{}{}.
\newblock
{\BBOQ}\APACrefatitle {Coordinating hundreds of cooperative, autonomous
  vehicles in warehouses} {Coordinating hundreds of cooperative, autonomous
  vehicles in warehouses}.{\BBCQ}
\newblock
\APACjournalVolNumPages{AI magazine}{29}{1}{9--9}.
\PrintBackRefs{\CurrentBib}

\bibitem [\protect \citeauthoryear {%
Yang%
, Jin%
\BCBL {}\ \BBA {} Duan%
}{%
Yang%
\ \protect \BOthers {.}}{%
{\protect \APACyear {2021}}%
}]{%
yangModellingAnalysisMultideep2021}
\APACinsertmetastar {%
yangModellingAnalysisMultideep2021}%
\begin{APACrefauthors}%
Yang, P.%
, Jin, G.%
\BCBL {}\ \BBA {} Duan, G.%
\end{APACrefauthors}%
\unskip\
\newblock
\APACrefYearMonthDay{2021}{}{}.
\newblock
{\BBOQ}\APACrefatitle {Modelling and analysis for multi-deep compact robotic
  mobile fulfilment system} {Modelling and analysis for multi-deep compact
  robotic mobile fulfilment system}.{\BBCQ}
\newblock
\APACjournalVolNumPages{International Journal of Production
  Research}{0}{0}{1--16}.
\PrintBackRefs{\CurrentBib}

\bibitem [\protect \citeauthoryear {%
Z{\"a}pfel%
\ \BBA {} Wasner%
}{%
Z{\"a}pfel%
\ \BBA {} Wasner%
}{%
{\protect \APACyear {2006}}%
}]{%
Zpfel2006WarehouseSI}
\APACinsertmetastar {%
Zpfel2006WarehouseSI}%
\begin{APACrefauthors}%
Z{\"a}pfel, G.%
\BCBT {}\ \BBA {} Wasner, M.%
\end{APACrefauthors}%
\unskip\
\newblock
\APACrefYearMonthDay{2006}{}{}.
\newblock
{\BBOQ}\APACrefatitle {Warehouse sequencing in the steel supply chain as a
  generalized job shop model} {Warehouse sequencing in the steel supply chain
  as a generalized job shop model}.{\BBCQ}
\newblock
\APACjournalVolNumPages{International Journal of Production
  Economics}{104}{}{482-501}.
\PrintBackRefs{\CurrentBib}

\bibitem [\protect \citeauthoryear {%
Zou%
, de Koster%
\BCBL {}\ \BBA {} Xu%
}{%
Zou%
\ \protect \BOthers {.}}{%
{\protect \APACyear {2016}}%
}]{%
zouEvaluatingDedicatedShared2016}
\APACinsertmetastar {%
zouEvaluatingDedicatedShared2016}%
\begin{APACrefauthors}%
Zou, B.%
, de Koster, M.%
\BCBL {}\ \BBA {} Xu, X.%
\end{APACrefauthors}%
\unskip\
\newblock
\APACrefYearMonthDay{2016}{}{}.
\newblock
{\BBOQ}\APACrefatitle {Evaluating dedicated and shared storage policies in
  robot-based compact storage and retrieval systems} {Evaluating dedicated and
  shared storage policies in robot-based compact storage and retrieval
  systems}.{\BBCQ}
\newblock
\APACjournalVolNumPages{ERIM Report Series Reference}{}{}{}.
\PrintBackRefs{\CurrentBib}

\end{thebibliography}

\appendix
\newpage
\section{Appendix}
\label{appendix}

\subsection{Detailed results for each parameter set}
\label{appendix_detailed_results}
This section contains the results for all parameter sets in tabular format.  Table~\ref{Tab:tier1_results_all} gives the results for single tier variants, Table~\ref{Tab:tier2_results_all} for the two tier variants, both having the same headers:
The first three columns contain the bay size, the access variant, and the fill level respectively.
Column four to six give the number of solved instances, the number of infeasible instances, and the number of instances for which a timeout occurred. 
The sum corresponds to the total number of instances which is ten in all cases.
The last three columns give the mean number of moves, the mean number of visited nodes during the A* search, and the mean total runtime over both phases of the algorithm over the number of solved instances (excluding the number of infeasible instances). 
If no solution could be found for any of the the instances the last three columns contain a dash (-).

\begin{longtable}{ M{1.5cm}M{1.25cm}M{1.25cm}M{1.25cm}M{1.25cm}M{1.25cm}M{1.25cm}M{1.25cm}M{1.25cm}}
\hline
Size (L\,x\,W\,x\,T) & Access variant & Fill level & No. solved & No. infeasible & No. timeout & Mean moves & Mean visited nodes & Mean runtime\\
\hline
\endhead
3x3x1 &    single &  40\% &        10 &           0 &        0 &   0.40 &           1.40 &     0.00 \\
3x3x1 &    single &  60\% &        10 &           0 &        0 &   1.80 &           3.10 &     0.01 \\
3x3x1 &    single &  80\% &         6 &           4 &        0 &   4.50 &           9.17 &     0.03 \\
3x3x1 &    corner &  40\% &        10 &           0 &        0 &   0.10 &           1.10 &     0.00 \\
3x3x1 &    corner &  60\% &        10 &           0 &        0 &   0.50 &           1.50 &     0.01 \\
3x3x1 &    corner &  80\% &        10 &           0 &        0 &   1.10 &           3.30 &     0.02 \\
3x3x1 &  opposite &  40\% &        10 &           0 &        0 &   0.00 &           1.00 &     0.00 \\
3x3x1 &  opposite &  60\% &        10 &           0 &        0 &   0.00 &           1.00 &     0.00 \\
3x3x1 &  opposite &  80\% &        10 &           0 &        0 &   0.30 &           1.30 &     0.00 \\
3x3x1 &     three &  40\% &        10 &           0 &        0 &   0.00 &           1.00 &     0.00 \\
3x3x1 &     three &  60\% &        10 &           0 &        0 &   0.00 &           1.00 &     0.00 \\
3x3x1 &     three &  80\% &        10 &           0 &        0 &   0.10 &           1.10 &     0.00 \\
3x3x1 &       four &  40\% &        10 &           0 &        0 &   0.00 &           1.00 &     0.00 \\
3x3x1 &       four &  60\% &        10 &           0 &        0 &   0.00 &           1.00 &     0.00 \\
3x3x1 &       four &  80\% &        10 &           0 &        0 &   0.00 &           1.00 &     0.00 \\
4x4x1 &    single &  40\% &        10 &           0 &        0 &   2.00 &           3.10 &     0.02 \\
4x4x1 &    single &  60\% &        10 &           0 &        0 &   5.60 &          17.30 &     0.10 \\
4x4x1 &    single &  80\% &         6 &           4 &        0 &   9.50 &         572.83 &     1.71 \\
4x4x1 &    corner &  40\% &        10 &           0 &        0 &   0.60 &           1.60 &     0.01 \\
4x4x1 &    corner &  60\% &        10 &           0 &        0 &   1.70 &           2.70 &     0.04 \\
4x4x1 &    corner &  80\% &        10 &           0 &        0 &   3.70 &          28.90 &     0.50 \\
4x4x1 &  opposite &  40\% &        10 &           0 &        0 &   0.30 &           1.30 &     0.01 \\
4x4x1 &  opposite &  60\% &        10 &           0 &        0 &   0.80 &           1.80 &     0.02 \\
4x4x1 &  opposite &  80\% &        10 &           0 &        0 &   1.80 &           2.80 &     0.03 \\
4x4x1 &     three &  40\% &        10 &           0 &        0 &   0.00 &           1.00 &     0.00 \\
4x4x1 &     three &  60\% &        10 &           0 &        0 &   0.70 &           1.70 &     0.03 \\
4x4x1 &     three &  80\% &        10 &           0 &        0 &   1.10 &           2.10 &     0.02 \\
4x4x1 &       four &  40\% &        10 &           0 &        0 &   0.00 &           1.00 &     0.00 \\
4x4x1 &       four &  60\% &        10 &           0 &        0 &   0.20 &           1.20 &     0.01 \\
4x4x1 &       four &  80\% &        10 &           0 &        0 &   0.70 &           1.70 &     0.02 \\
5x5x1 &    single &  40\% &        10 &           0 &        0 &   3.40 &           4.40 &     0.06 \\
5x5x1 &    single &  60\% &        10 &           0 &        0 &   8.10 &          25.30 &     0.29 \\
5x5x1 &    single &  80\% &        10 &           0 &        0 &  16.80 &       13700.10 &    93.17 \\
5x5x1 &    corner &  40\% &        10 &           0 &        0 &   1.60 &           2.60 &     0.09 \\
5x5x1 &    corner &  60\% &        10 &           0 &        0 &   3.00 &           4.60 &     0.20 \\
5x5x1 &    corner &  80\% &        10 &           0 &        0 &   6.80 &         104.50 &     4.78 \\
5x5x1 &  opposite &  40\% &        10 &           0 &        0 &   0.90 &           1.90 &     0.06 \\
5x5x1 &  opposite &  60\% &        10 &           0 &        0 &   2.30 &           3.30 &     0.17 \\
5x5x1 &  opposite &  80\% &        10 &           0 &        0 &   4.30 &           5.30 &     0.22 \\
5x5x1 &     three &  40\% &        10 &           0 &        0 &   0.90 &           1.90 &     0.09 \\
5x5x1 &     three &  60\% &        10 &           0 &        0 &   1.30 &           2.30 &     0.12 \\
5x5x1 &     three &  80\% &        10 &           0 &        0 &   1.70 &           2.70 &     0.11 \\
5x5x1 &       four &  40\% &        10 &           0 &        0 &   0.10 &           1.10 &     0.02 \\
5x5x1 &       four &  60\% &        10 &           0 &        0 &   0.80 &           1.80 &     0.10 \\
5x5x1 &       four &  80\% &        10 &           0 &        0 &   1.30 &           2.30 &     0.12 \\
6x6x1 &    single &  40\% &        10 &           0 &        0 &   7.40 &          17.20 &     0.38 \\
6x6x1 &    single &  60\% &        10 &           0 &        0 &  14.20 &         495.70 &    10.79 \\
6x6x1 &    single &  80\% &         4 &           0 &        6 &  23.25 &       19879.75 &   297.37 \\
6x6x1 &    corner &  40\% &        10 &           0 &        0 &   3.30 &           4.30 &     0.38 \\
6x6x1 &    corner &  60\% &        10 &           0 &        0 &   6.40 &           7.40 &     0.77 \\
6x6x1 &    corner &  80\% &         9 &           0 &        1 &  11.00 &        4536.78 &   368.82 \\
6x6x1 &  opposite &  40\% &        10 &           0 &        0 &   1.30 &           2.30 &     0.19 \\
6x6x1 &  opposite &  60\% &        10 &           0 &        0 &   3.10 &           4.10 &     0.52 \\
6x6x1 &  opposite &  80\% &        10 &           0 &        0 &   7.70 &          63.40 &     6.18 \\
6x6x1 &     three &  40\% &        10 &           0 &        0 &   1.10 &           2.10 &     0.25 \\
6x6x1 &     three &  60\% &        10 &           0 &        0 &   3.30 &           4.30 &     0.75 \\
6x6x1 &     three &  80\% &        10 &           0 &        0 &   4.50 &           5.50 &     0.77 \\
6x6x1 &       four &  40\% &        10 &           0 &        0 &   0.30 &           1.30 &     0.09 \\
6x6x1 &       four &  60\% &        10 &           0 &        0 &   1.20 &           2.20 &     0.41 \\
6x6x1 &       four &  80\% &        10 &           0 &        0 &   2.70 &           3.70 &     0.53 \\
7x7x1 &    single &  40\% &        10 &           0 &        0 &  10.80 &          53.50 &     2.27 \\
7x7x1 &    single &  60\% &        10 &           0 &        0 &  18.50 &        2600.10 &   103.96 \\
7x7x1 &    single &  80\% &         0 &           0 &       10 &    - &            - &      - \\
7x7x1 &    corner &  40\% &        10 &           0 &        0 &   7.20 &           8.20 &     1.76 \\
7x7x1 &    corner &  60\% &        10 &           0 &        0 &  10.60 &          11.60 &     2.75 \\
7x7x1 &    corner &  80\% &         5 &           0 &        5 &  16.20 &        4683.60 &  1107.87 \\
7x7x1 &  opposite &  40\% &        10 &           0 &        0 &   1.80 &           2.80 &     0.65 \\
7x7x1 &  opposite &  60\% &        10 &           0 &        0 &   6.00 &           9.90 &     3.13 \\
7x7x1 &  opposite &  80\% &        10 &           0 &        0 &  11.60 &         122.90 &    35.53 \\
7x7x1 &     three &  40\% &        10 &           0 &        0 &   2.00 &           3.00 &     1.02 \\
7x7x1 &     three &  60\% &        10 &           0 &        0 &   4.40 &           5.50 &     2.16 \\
7x7x1 &     three &  80\% &        10 &           0 &        0 &   8.60 &          14.30 &     6.33 \\
7x7x1 &       four &  40\% &        10 &           0 &        0 &   1.30 &           2.30 &     0.82 \\
7x7x1 &       four &  60\% &        10 &           0 &        0 &   2.50 &           3.50 &     1.66 \\
7x7x1 &       four &  80\% &        10 &           0 &        0 &   5.60 &           6.60 &     2.85 \\
8x8x1 &    single &  40\% &        10 &           0 &        0 &  13.40 &        1916.70 &   137.25 \\
8x8x1 &    single &  60\% &         7 &           0 &        3 &  25.57 &       12340.14 &   823.97 \\
8x8x1 &    single &  80\% &         0 &           0 &       10 &    - &            - &      - \\
8x8x1 &    corner &  40\% &        10 &           0 &        0 &   8.10 &           9.10 &     3.49 \\
8x8x1 &    corner &  60\% &        10 &           0 &        0 &  14.20 &          28.50 &    12.57 \\
8x8x1 &    corner &  80\% &         2 &           0 &        8 &  22.00 &          82.50 &    21.61 \\
8x8x1 &  opposite &  40\% &        10 &           0 &        0 &   4.60 &           5.60 &     2.72 \\
8x8x1 &  opposite &  60\% &        10 &           0 &        0 &   9.70 &          10.70 &     6.06 \\
8x8x1 &  opposite &  80\% &         8 &           0 &        2 &  16.25 &         257.38 &    76.50 \\
8x8x1 &     three &  40\% &        10 &           0 &        0 &   4.30 &           5.30 &     3.85 \\
8x8x1 &     three &  60\% &        10 &           0 &        0 &   6.70 &           7.70 &     5.94 \\
8x8x1 &     three &  80\% &         8 &           0 &        2 &  13.75 &         510.75 &   306.57 \\
8x8x1 &       four &  40\% &        10 &           0 &        0 &   2.70 &           3.70 &     3.52 \\
8x8x1 &       four &  60\% &        10 &           0 &        0 &   5.20 &           6.20 &     6.79 \\
8x8x1 &       four &  80\% &        10 &           0 &        0 &   9.20 &          10.20 &     7.97 \\
9x9x1 &    single &  40\% &         9 &           0 &        1 &  17.89 &         205.33 &    42.39 \\
9x9x1 &    single &  60\% &         3 &           0 &        7 &  30.33 &        1540.67 &   154.91 \\
9x9x1 &    single &  80\% &         0 &           0 &       10 &    - &            - &      - \\
9x9x1 &    corner &  40\% &        10 &           0 &        0 &  10.50 &          11.50 &     7.38 \\
9x9x1 &    corner &  60\% &         9 &           0 &        1 &  21.56 &          22.56 &    16.15 \\
9x9x1 &    corner &  80\% &         1 &           0 &        9 &  36.00 &        3515.00 &  1529.92 \\
9x9x1 &  opposite &  40\% &        10 &           0 &        0 &   5.60 &           6.60 &     5.33 \\
9x9x1 &  opposite &  60\% &        10 &           0 &        0 &  14.00 &          15.20 &    13.94 \\
9x9x1 &  opposite &  80\% &         2 &           0 &        8 &  22.00 &          43.00 &    24.12 \\
9x9x1 &     three &  40\% &        10 &           0 &        0 &   5.10 &           6.10 &     7.58 \\
9x9x1 &     three &  60\% &        10 &           0 &        0 &  12.50 &          13.50 &    21.43 \\
9x9x1 &     three &  80\% &         8 &           0 &        2 &  18.38 &          82.62 &   133.51 \\
9x9x1 &       four &  40\% &        10 &           0 &        0 &   4.00 &           5.00 &     8.59 \\
9x9x1 &       four &  60\% &        10 &           0 &        0 &   8.50 &           9.50 &    22.32 \\
9x9x1 &       four &  80\% &        10 &           0 &        0 &  14.60 &          15.60 &    19.42 \\
10x10x1 &    single &  40\% &         7 &           0 &        3 &  22.14 &         240.57 &    39.09 \\
10x10x1 &    single &  60\% &         0 &           0 &       10 &    - &            - &      - \\
10x10x1 &    single &  80\% &         0 &           0 &       10 &    - &            - &      - \\
10x10x1 &    corner &  40\% &         10 &           0 &        0 &  16.20 &          17.20 &    16.78 \\
10x10x1 &    corner &  60\% &         8 &           0 &        2 &  30.00 &          41.62 &    43.20 \\
10x10x1 &    corner &  80\% &         0 &           0 &       10 &    - &            - &      - \\
10x10x1 &  opposite &  40\% &        10 &           0 &        0 &   8.70 &           9.70 &    13.22 \\
10x10x1 &  opposite &  60\% &         9 &           0 &        1 &  18.89 &          21.44 &    31.16 \\
10x10x1 &  opposite &  80\% &         0 &           0 &       10 &    - &            - &      - \\
10x10x1 &     three &  40\% &        10 &           0 &        0 &   7.30 &           8.30 &    16.85 \\
10x10x1 &     three &  60\% &        10 &           0 &        0 &  17.70 &          18.70 &    41.91 \\
10x10x1 &     three &  80\% &         6 &           0 &        4 &  28.50 &         334.67 &   441.85 \\
10x10x1 &       four &  40\% &        10 &           0 &        0 &   6.40 &           7.40 &    22.71 \\
10x10x1 &       four &  60\% &        10 &           0 &        0 &  11.20 &          12.20 &    39.44 \\
10x10x1 &       four &  80\% &         7 &           0 &        3 &  18.86 &          31.14 &    65.20 \\
\hline
\caption{Results for a single tier rounded to the second decimal}
\label{Tab:tier1_results_all}
\end{longtable}

\begin{table}[]
\centering
\begin{tabular}{ M{1.5cm}M{1.25cm}M{1.25cm}M{1.25cm}M{1.25cm}M{1.25cm}M{1.25cm}M{1.25cm}M{1.25cm}}
\hline
Size (L\,x\,W\,x\,T) & Access variant & Fill level & No. solved & No. infeasible & No. timeout & Mean moves & Mean visited nodes & Mean runtime\\
\hline
3x3x2 &    single &  40\% &        10 &           0 &        0 &   4.80 &          10.20 &     0.04 \\
3x3x2 &    single &  60\% &        10 &           0 &        0 &   9.80 &         296.70 &     0.64 \\
3x3x2 &    single &  80\% &         0 &          10 &        0 &    - &            - &      - \\
3x3x2 &    corner &  40\% &        10 &           0 &        0 &   1.80 &           2.80 &     0.02 \\
3x3x2 &    corner &  60\% &        10 &           0 &        0 &   4.70 &           6.90 &     0.07 \\
3x3x2 &    corner &  80\% &         5 &           0 &        5 &   9.00 &         144.60 &     1.51 \\
3x3x2 &  opposite &  40\% &        10 &           0 &        0 &   1.20 &           2.20 &     0.02 \\
3x3x2 &  opposite &  60\% &        10 &           0 &        0 &   3.30 &           4.70 &     0.06 \\
3x3x2 &  opposite &  80\% &        10 &           0 &        0 &   4.60 &          14.90 &     0.20 \\
3x3x2 &     three &  40\% &        10 &           0 &        0 &   0.90 &           1.90 &     0.02 \\
3x3x2 &     three &  60\% &        10 &           0 &        0 &   2.20 &           3.20 &     0.04 \\
3x3x2 &     three &  80\% &        10 &           0 &        0 &   5.20 &          21.10 &     0.32 \\
3x3x2 &       four &  40\% &        10 &           0 &        0 &   1.10 &           2.10 &     0.03 \\
3x3x2 &       four &  60\% &        10 &           0 &        0 &   1.70 &           2.70 &     0.04 \\
3x3x2 &       four &  80\% &        10 &           0 &        0 &   2.80 &           3.80 &     0.06 \\
4x4x2 &    single &  40\% &        10 &           0 &        0 &   8.40 &          93.80 &     0.65 \\
4x4x2 &    single &  60\% &        10 &           0 &        0 &  16.50 &       31331.60 &   183.02 \\
4x4x2 &    single &  80\% &         0 &           0 &       10 &    - &            - &      - \\
4x4x2 &    corner &  40\% &        10 &           0 &        0 &   5.30 &           6.50 &     0.17 \\
4x4x2 &    corner &  60\% &        10 &           0 &        0 &   9.60 &          69.30 &     3.12 \\
4x4x2 &    corner &  80\% &         5 &           0 &        5 &  14.00 &       16817.60 &   553.42 \\
4x4x2 &  opposite &  40\% &        10 &           0 &        0 &   3.00 &           4.00 &     0.14 \\
4x4x2 &  opposite &  60\% &        10 &           0 &        0 &   6.80 &           8.10 &     0.37 \\
4x4x2 &  opposite &  80\% &        10 &           0 &        0 &  13.00 &        4141.80 &   137.62 \\
4x4x2 &     three &  40\% &        10 &           0 &        0 &   3.00 &           4.00 &     0.20 \\
4x4x2 &     three &  60\% &        10 &           0 &        0 &   4.90 &           5.90 &     0.33 \\
4x4x2 &     three &  80\% &        10 &           0 &        0 &  10.30 &        3987.20 &   204.97 \\
4x4x2 &       four &  40\% &        10 &           0 &        0 &   2.50 &           3.50 &     0.22 \\
4x4x2 &       four &  60\% &        10 &           0 &        0 &   4.90 &           5.90 &     0.42 \\
4x4x2 &       four &  80\% &        10 &           0 &        0 &   7.30 &          60.90 &     4.16 \\
5x5x2 &    single &  40\% &        10 &           0 &        0 &  13.50 &        3028.70 &    46.02 \\
5x5x2 &    single &  60\% &         5 &           0 &        5 &  24.00 &       47003.40 &   652.39 \\
5x5x2 &    single &  80\% &         0 &           0 &       10 &    - &            - &      - \\
5x5x2 &    corner &  40\% &        10 &           0 &        0 &   9.40 &          11.30 &     0.79 \\
5x5x2 &    corner &  60\% &         7 &           0 &        3 &  15.00 &        3305.43 &   495.75 \\
5x5x2 &    corner &  80\% &         0 &           0 &       10 &    - &            - &      - \\
5x5x2 &  opposite &  40\% &        10 &           0 &        0 &   6.30 &           7.30 &     0.71 \\
5x5x2 &  opposite &  60\% &        10 &           0 &        0 &  11.10 &         221.10 &    32.85 \\
5x5x2 &  opposite &  80\% &         2 &           0 &        8 &  21.00 &       13519.50 &  1478.46 \\
5x5x2 &     three &  40\% &        10 &           0 &        0 &   6.00 &           7.00 &     1.00 \\
5x5x2 &     three &  60\% &        10 &           0 &        0 &  10.00 &          11.00 &     3.09 \\
5x5x2 &     three &  80\% &         4 &           0 &        6 &  16.25 &        6282.00 &   751.99 \\
5x5x2 &       four &  40\% &        10 &           0 &        0 &   4.90 &           5.90 &     1.24 \\
5x5x2 &       four &  60\% &        10 &           0 &        0 &   7.70 &           8.70 &     3.54 \\
5x5x2 &       four &  80\% &         7 &           0 &        3 &  13.14 &         629.71 &   130.59 \\
\hline
\end{tabular}
\caption{\label{Tab:tier2_results_all} Results for two tiers rounded to the second decimal}
\end{table}

\newpage
\subsection{Box plots for the number of moves for all sizes}
\label{appendix_moves}

\begin{figure}[htp]
    \centering
    \includegraphics[width=10cm]{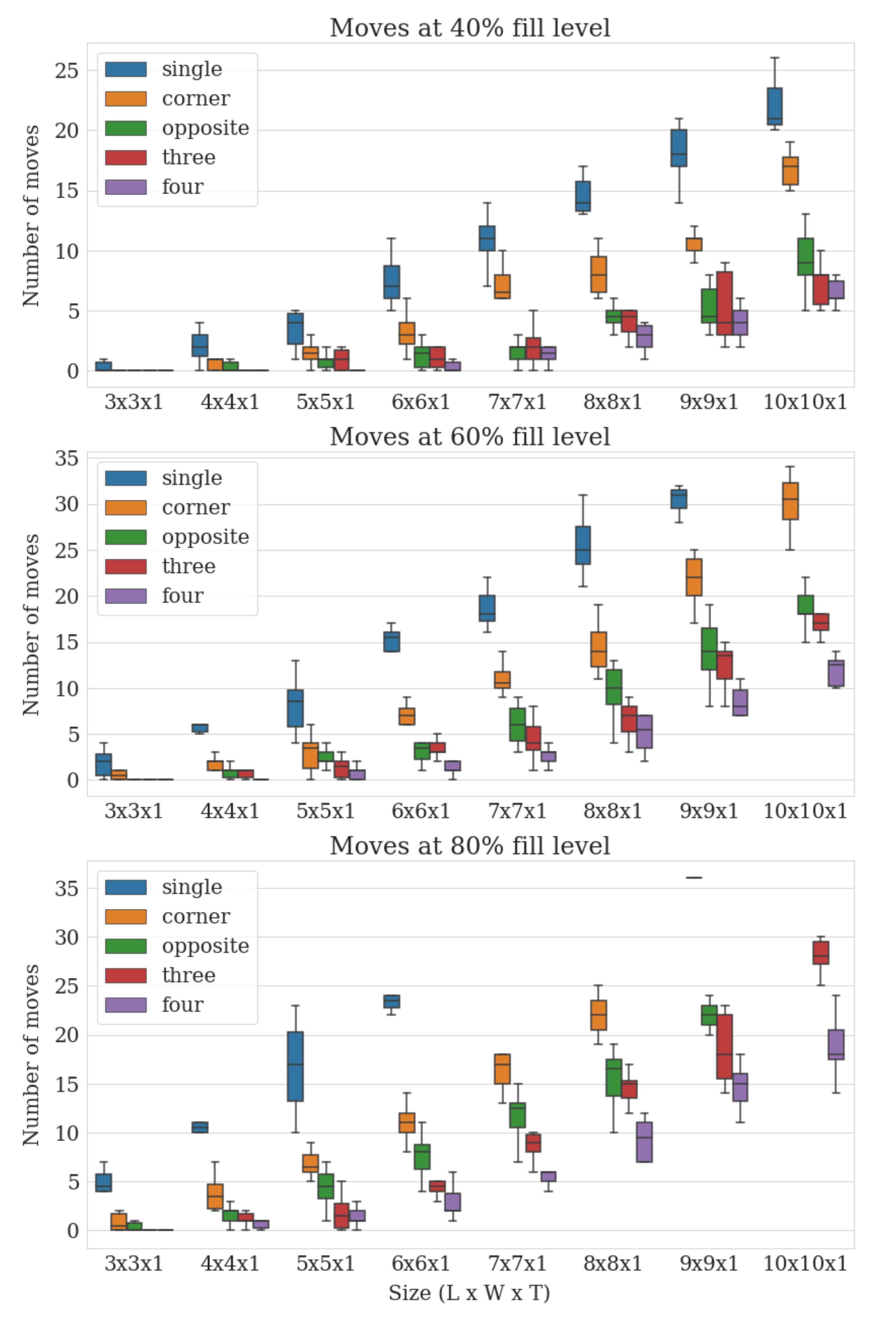}
    \caption{Box plots (without outliers) for a single tier showing the number of moves for each access variant for all sizes and each fill level. This is based on the solved instances shown in Figure \ref{fig:instances_runtime}.}
    \label{fig:moves_boxbplot_tier1}
\end{figure}

\begin{figure}[htp]
    \centering
    \includegraphics[width=5cm]{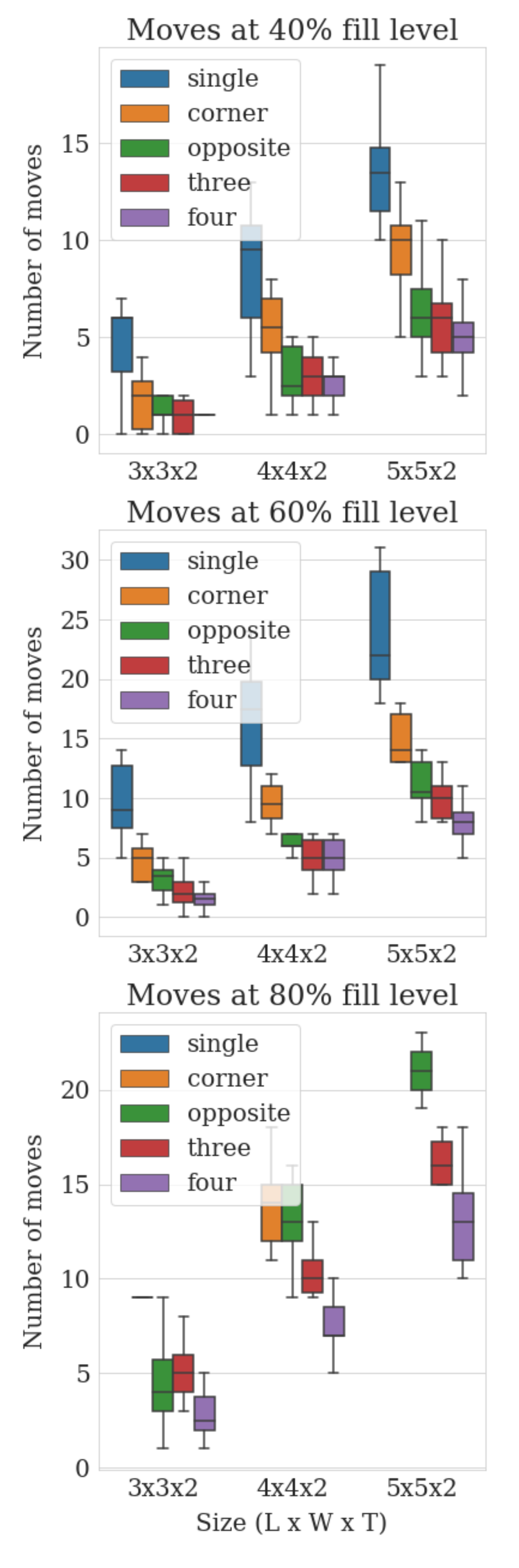}
    \caption{Boxplots (without outliers) for two tiers showing the number of moves for each access variant for all sizes and each fill level. This is based on the solved instances shown in Figure \ref{fig:instances_runtime_two_tiers}.}
    \label{fig:moves_boxbplot_tier2}
\end{figure}

This section contains boxplots for the number of moves for differentiated by setup.
Figure~\ref{fig:moves_boxbplot_tier1} contains the results for the single tier instances, Figure~\ref{fig:moves_boxbplot_tier2} the results for the two tier instances.
The results are differentiated by the size of the bays on the x-axis (L\,x\,W\,x\,T), the fill level in the row, and by the considered access directions using colors.

\end{document}